# Comparison of optical spectra between asteroids Ryugu and Bennu: I. Cross calibration between Hayabusa2/ONC-T and OSIRIS-REx/MapCam

K. Yumoto[a, *], E. Tatsumi[b], T. Kouyama[c], D. R. Golish[d], Y. Cho[a], T. Morota[a], S. Kameda[e, b], H. Sato[b], B. Rizk[d], D. N. DellaGiustina[d], Y. Yokota[b], H. Suzuki[f], J. de León[g, h], H. Campins[i], J. Licandro[g, h], M. Popescu[j], J. L. Rizos[k, l], R. Honda[m, †], M. Yamada[n], N. Sakatani[b], C. Honda[o], M. Matsuoka[p], M. Hayakawa[b], H. Sawada[b], K. Ogawa[q], Y. Yamamoto[b], D. S. Lauretta[d], S. Sugita[a, n, r]

[a]Department of Earth and Planetary Science, The University of Tokyo, Bunkyo, Tokyo, Japan.
[b]Institute of Space and Astronautical Science, Japan Aerospace Exploration Agency, Sagamihara, Kanagawa, Japan.
[c]Artificial Intelligence Research Center, National Institute of Advanced Industrial Science and Technology, Koto, Tokyo, Japan.
[d]Lunar and Planetary Laboratory, University of Arizona, Tucson, AZ, USA
[e]Department of Physics, Rikkyo University, Toshima, Tokyo, Japan.
[f]Department of Physics, Meiji University, Kawasaki, Kanagawa, Japan.
[g]Instituto de Astrofísica de Canarias (IAC), University of La Laguna, La Laguna, Tenerife, Spain.
[h]Department of Astrophysics, University of La Laguna, La Laguna, Tenerife, Spain.
[i]Department of Physics, University of Central Florida, Orlando, FL, USA.
[j]Astronomical Institute of the Romanian Academy, Bucharest, Romania.
[k]Instituto de Astrofísica de Andalcía (IAA) – CSIC, Granada, Andalcía, Spain.
[l]University of Maryland, College Park, MD, USA.
[m]Center for Data Science, Ehime University, Matsuyama, Ehime, Japan.
[n]Planetary Exploration Research Center (PERC), Chiba Institute of Technology, Narashino, Chiba, Japan.
[o]The University of Aizu, Aizu-Wakamatsu, Fukushima, Japan.
[p]Geological Survey of Japan (GSJ), National Institute of Advanced Industrial Science and Technology, Tsukuba, Ibaraki, Japan
[q]JAXA Space Exploration Center, JAXA, Sagamihara, Japan.
[r]Research Center of Early Universe, Graduate School of Science, The University of Tokyo, Tokyo, Japan.
†Deceased

Corresponding author: Koki Yumoto (kyyumoto@gmail.com)

**Highlights**

- We cross calibrated the two imagers using the Moon as the common standard.
- We reduced uncertainty in the *v*-band reflectance of Ryugu to Bennu from 15% to <2%.
- Future studies can apply our results by multiplying a constant to the Bennu data.
- The cross-calibrated geometric albedo is 4.1 ± 0.1% for Ryugu and 4.9 ± 0.1% for Bennu.
- The results were validated against observations by the OSIRIS-REx visible and infrared spectrometer.

**Abstract**

Asteroids (162173) Ryugu and (101955) Bennu observed by Hayabusa2 and Origins, Spectral Interpretation, Resource Identification, and Security-Regolith Explorer (OSIRIS-REx) share many global properties, but high-spatial-resolution spectral observations by the telescopic Optical Navigation Camera (ONC-T) and MapCam detected subtle but significant differences (e.g., opposite space weathering trends), which may reflect differences in their origin and evolution. Comparing these differences on the same absolute scale is necessary for understanding their causes and obtaining implications for C-complex asteroids. However, ONC-T and MapCam have a large imager-to-imager systematic error of up to 15% caused by the difference in radiometric calibration targets. To resolve this problem, we cross calibrated albedo and color data between the two instruments using the Moon as the common calibration standard. The images of the Moon taken by ONC-T and MapCam were compared with those simulated using photometry models developed from lunar orbiter data. Our results show that the cross-calibrated reflectance of Ryugu and Bennu can be obtained by upscaling the pre-cross-calibrated reflectance of Bennu by 13.3 ± 1.6% at *b* band, 13.2 ± 1.5% at *v* band, 13.6 ± 1.7% at *w* band, and 14.8 ± 1.8% at *x* band, while those for Ryugu are kept the same. These factors compensate for the imager-to-imager bias caused by differences in targets used for radiometric calibration and solar irradiance models used for data reduction. Need for such large upscaling underscore the importance of using the cross-calibrated data for accurately comparing the Ryugu and Bennu data. The uncertainty in these factors show that the reflectance of Ryugu and Bennu can be compared with <2% accuracy after applying our results. By applying our cross calibration, the geometric albedo of Bennu became consistent with those observed by ground-based telescopes and the OSIRIS-REx Visible and InfraRed Spectrometer (OVIRS). Our result can be simply applied by multiplying a constant to the publicly available data and enables accurate comparison of the optical spectra of Ryugu and Bennu in future studies.

## 1. Introduction

Analyses of the compositional structure of the asteroid belt have long relied on optical spectra collected through ground-based telescopes. Albedo was one of the primal observables in the early 1970's, and its bimodal distribution has led to the well-known distinction between "stony" and "carbonaceous"-type asteroids (Zellner, 1973). During the following decades, spectroscopic observations using charge-coupled devices (CCDs) grew rapidly and revealed the wide spectral variation in the asteroid belt (Zellner et al., 1985; Xu et al., 1995; Bus & Binzel, 2002a; Masiero et al., 2011; Ivezic et al., 2020). Asteroid taxonomy based on their optical spectra (Tholen, 1984; Bus & Binzel, 2002b; Carvano et al., 2010) is widely used today and have driven the understanding of compositional variation associated with orbital parameters and asteroid size (DeMeo & Carry, 2014). The spectra of meteoritic samples were also extensively measured in laboratories and imparted insights into the mineralogy and physical properties of asteroid surfaces (e.g., Hiroi et al., 1996; Cloutis et al., 2011). Because the spectra in the wavelength range compatible with CCDs (0.4–1 μm) are still the largest dataset for asteroid characterization, they play a key role in putting individual asteroids in the context of the entire asteroid population.

The spectral variation of C-complex asteroids spans over a wide range, which probably reflect heterogeneity among their parent bodies and space weathering effects (Fornasier et al., 2016; Kaluna et al., 2016). Comprehensively understanding the cause of spectral variation solely from telescopic data has been challenging, as multiple processes can produce similar effects on the optical spectrum. While spectral measurements of meteorites can aid in addressing this issue, it is important to note that C-complex asteroids do not always have a reliable spectral analogue in meteorite samples (Britt et al., 1992). Proximity spectral observations with spacecraft can help solve this problem by resolving the spectra among different geologic features on the asteroid surface. For example, spectral observations of (1) Ceres by Dawn's Framing Camera found spectrally blue materials within the floor and ejecta of 10-km-sized fresh impact craters (Schmedemann et al., 2016; Stephan et al., 2017). This bluing may be caused by the impact-induced mixing of phyllosilicates with subsurface ice (Schröder et al., 2021), paving the way for understanding the spectral variation among ice-bearing large (>10 km) asteroids in the middle to outer main belt (Schörghofer & Hsieh, 2018).

Proximity observations of the near-Earth C-type asteroids (162173) Ryugu by Hayabusa2 and (101955) Bennu by Origins, Spectral Interpretation, Resource Identification, and Security-Regolith Explorer (OSIRIS-REx) may provide important constraints for understanding the spectral variation among smaller (≲1 km) C-complex asteroids. These two missions carried multi-band CCD imagers: the telescopic Optical Navigation Camera (ONC-T) for Hayabusa2 and the MapCam medium-field imager for OSIRIS-REx (see Table 1 for their specifications). These imagers globally observed the extended visible spectra (0.48–0.85 μm) of the asteroids with sub-meter spatial resolutions. Observing C-type asteroids with such an unprecedented spatial resolution opens a new horizon for spectral investigation. For instance, we can evaluate the heterogeneity in the building fragments of these asteroids by resolving the spectra of individual boulders.

Ryugu and Bennu share many spectral and geologic properties, such as extremely low albedo, similar optical spectral types (Cb and B), high boulder abundance, and lack of regolith ponds (Lauretta et al., 2019; Sugita et al., 2019). However, detailed investigation of their optical spectra by ONC-T and MapCam revealed many qualitative dissimilarities, suggesting differences in the origin and evolution of these asteroids. For instance, Bennu surface exhibits a larger range of albedo variation (DellaGiustina et al., 2019; 2020), spectral trends of space weathering appear to be opposite between Ryugu and Bennu (Sugita et al., 2019; Morota et al., 2020; DellaGiustina et al., 2020; Tatsumi et al., 2021c; Lauretta et al., 2022), and the spectra of some bright exogenic boulders are consistent with different known types of meteorites/asteroids (Tatsumi et al., 2021a; 2021b; DellaGiustina et al., 2021).

Placing ONC-T and MapCam data in the same absolute scale is the next step to quantitatively analyze these findings and detect more subtle differences/similarities. For instance, comparing the global average albedo between Ryugu and Bennu along with their range of heterogeneity has an implication for their parent-body material. Comparing the space weathering trends on the same scale is also important to understand whether the spectra of fresh materials on Ryugu and Bennu are similar (see Yumoto et al., 2024 for details). Comparing the spectra of exogenic boulders may tell us if Ryugu and Bennu experienced impact-induced mixing with similar materials (Tatsumi et al., 2021b).

Such direct imager-to-imager comparison is potentially possible because these imagers are equipped with four narrow-band filters centered at similar wavelengths (Table 1). However, such a precise analysis requires reduction in the systematic calibration difference between ONC-T and MapCam data. Such systematic difference mainly arises from the fact that these two imagers used different standard targets for their radiometric calibration; ONC-T is calibrated to stars while MapCam is calibrated to the Moon. Such difference in calibration targets originates from fundamental differences in their hardware (section 3). In this study, we call this systematic multiplicative difference between ONC-T and MapCam data as the imager-to-imager bias for conciseness. This imager-to-imager bias in absolute radiometric responsivities can be as high as 15% (section 3). By contrast, the global average reflectance of Ryugu and Bennu differ by only 6% in the pre-cross-calibrated data (DellaGiustina et al., 2020). Thus, we cannot even decisively conclude which asteroid has a higher reflectance due to the large imager-to-imager bias; reducing the bias down to 2–3% is required for a definite conclusion.

To resolve this problem, we conducted a cross calibration between ONC-T and MapCam using the Moon as the common calibration target. We only used the cross-calibration results for correcting the imager-to-imager bias and chose to rely on the star-based calibration of ONC-T for the absolute calibration (section 4.1).

In the following, we first discuss the comparability between ONC-T and MapCam in section 2. In section 3, we identify the cause for the imager-to-imager bias. In section 4, we discuss the methods and introduce the correction factors for obtaining the cross-calibrated reflectance. In section 5, we present the resulting values and uncertainties of these factors. In section 6, we summarize how our results can be applied to the publicly available pre-cross-calibrated data of Ryugu and Bennu. Then, we show the albedos and spectra of Ryugu and Bennu after applying our results in section 7 before we conclude in section 8. Comparative analyses of space weathering effects on Ryugu and Bennu using the cross-calibrated images will be presented in our companion paper (Yumoto et al., 2024).

**Table 1.** Specifications of ONC-T and MapCam.

| Imager | CCD model (manufacturer)[1, 2]† | Mean plate scale (μrad/pix)[2, 3] | Bands[4, 5] | | |
|---|---|---|---|---|---|
| | | | Band name | Band center (nm) | Band widths (nm) |
| ONC-T | CCD4720AIMO (Teledyne e2v) | 107 | *ul* | 398 | 36 |
| | | | *b* | 480 | 27 |
| | | | *v* | 549 | 31 |
| | | | *Na* | 590 | 12 |
| | | | *w* | 700 | 29 |
| | | | *x* | 857 | 42 |
| | | | *p* | 945 | 56 |
| MapCam | Custom-built (Teledyne DALSA Custom division) | 68 | *b'* | 473 | 61 |
| | | | *v* | 550 | 57 |
| | | | *w* | 698 | 60 |
| | | | *x* | 847 | 78 |

[1] Kameda et al. (2017) [2] Rizk et al. (2018) [3] Suzuki et al. (2018) [4] Tatsumi et al. (2019) [5] Golish et al. (2020a).

† See Fig. 1 for the detector layouts.

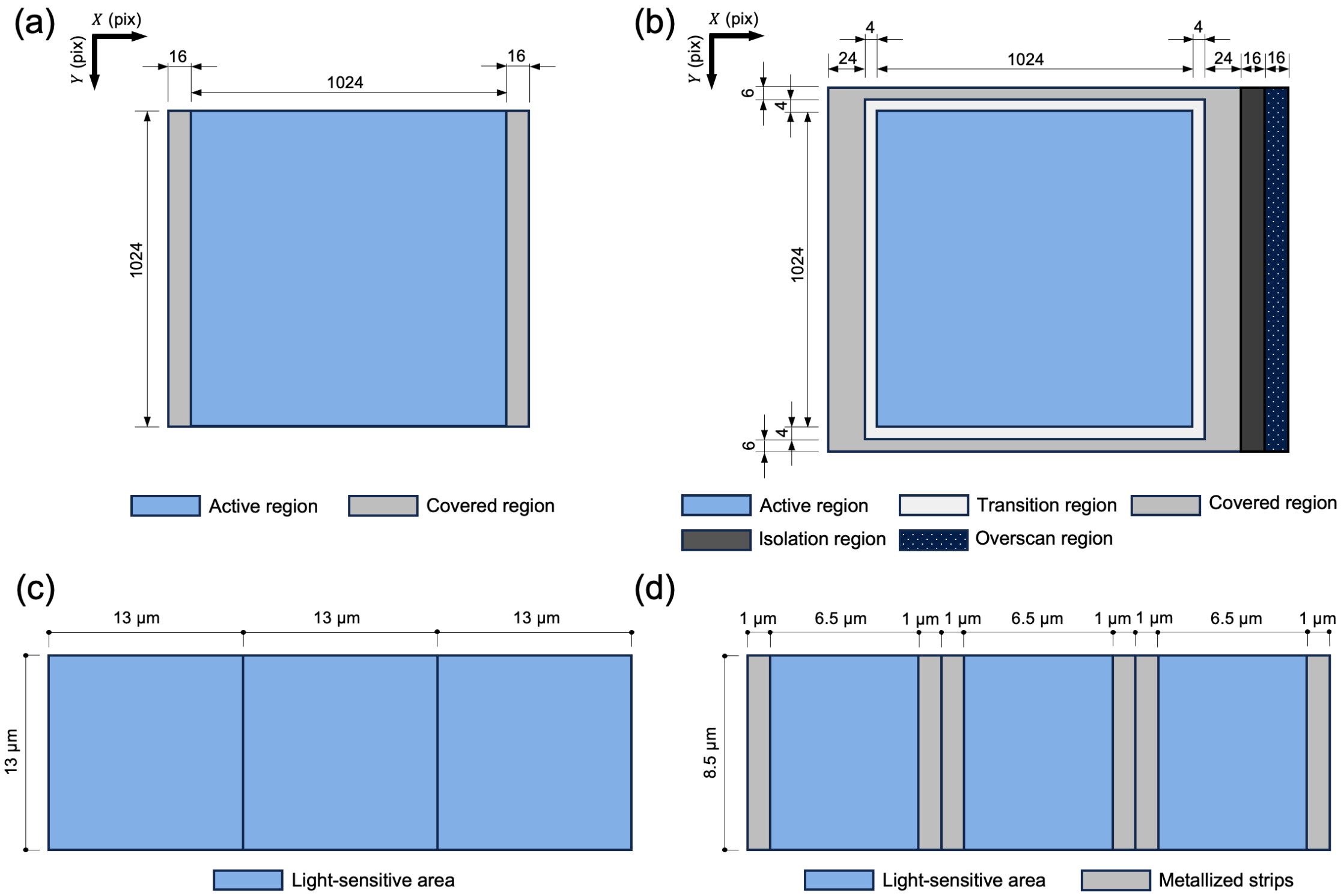

**Figure 1.** Layouts of CCDs for ONC-T and MapCam. The configurations of detectors in the horizontal ($X$) and vertical ($Y$) directions for ONC-T and MapCam are illustrated in **(a)** and **(b)**, respectively. The imaging process utilizes *active regions* of 1024 × 1024 pixels, with masked pixels in the *covered regions* for dark reference acquisition. The *isolation* and *overscan* regions of MapCam include virtual pixels (pixels that are read out but not associated with physical detectors) for the evaluation of bias level. Dimensions of each detector and the sub-pixel structure for ONC-T and MapCam are depicted in **(c)** and **(d)**, respectively. ONC-T detectors are fully light-sensitive, while 24% of each MapCam detector is partially masked with metallized strips positioned on anti-blooming barriers and drains.

## 2. Comparability between ONC-T and MapCam

In this section, we examine the spectral responsivity of each filter on ONC-T and MapCam to show that the spectra observed by the two imagers can be accurately compared despite the difference in their hardware.

The spectral responsivity ($\phi_n$) of the optical system at different wavelengths ($\lambda$) depends on the transmission efficiency of the optical components and the quantum efficiency of the detector. The subscript $n$ indicates the pertinent bands ($n = b$, $v$, $w$, $x$ band). Since different optical components and detectors are used for ONC-T and MapCam, $\phi_n$ of the shared four filters have differences in its central wavelength and width at ~10 nm scales (Fig. 2). Such differences in $\phi_n$ may limit the direct comparison of the band spectrum observed by the two imagers.

To quantitatively evaluate this limitation, we simulated the band spectra ($R_n$) observed by the two imagers using the same reference spectrum ($R_{ref}$) as follows:

$$R_n = \frac{\int R_{ref}(\lambda)/R_{ref}(550\ \text{nm})\, \lambda \phi_n(\lambda)\, d\lambda}{\int \lambda \phi_n(\lambda)\, d\lambda}. \quad (1)$$

The spectra of Ryugu and Bennu observed by ground-based telescopes were used as $R_{ref}$. For Ryugu, we used the spectra observed by the Magellan 6.5-m telescope (Moskovitz et al., 2013) and the 10.4-m Gran Telescopio Canarias (Tatsumi et al., 2022). For Bennu, we used the spectra observed by the McDonald Observatory 2.1-m telescope (Clark et al., 2011) and the University of Arizona Kuiper 1.54-m telescope (Hergenrother et al., 2013). To broaden the investigation, we also used the spectra of C-complex asteroids observed by SMASS II ($N$ =396; Bus & Binzel, 2002a) for $R_{ref}$. The data for $\phi_n$ and $R_{ref}$ were resampled to 1-nm resolution prior to calculating the integration in equation 1.

The band spectra simulated for MapCam ($R_n^{\text{MapCam}}$) and ONC-T ($R_n^{\text{ONC}}$) match well with each other. The values of $R_n^{\text{MapCam}}/R_n^{\text{ONC}}$-1 for each $R_{ref}$ (Fig. 3 and Table2) show that the band spectra observed by the two imagers agree within <1%, demonstrating that the spectra observed by the two different imagers can be ideally compared with <1% accuracy.

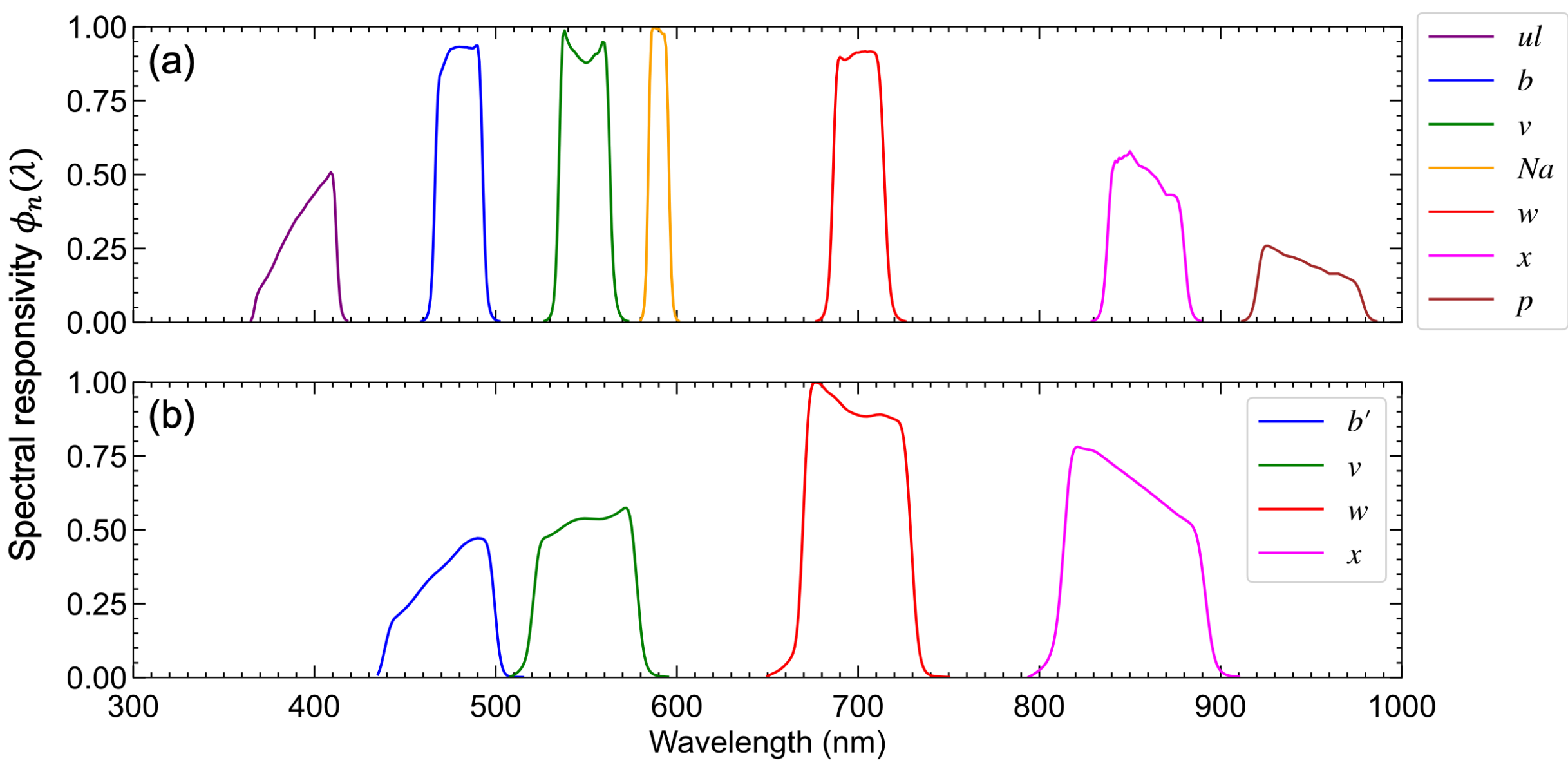

**Figure 2.** Spectral responsivity ($\boldsymbol{\phi_n}$) of each filter onboard **(a)** ONC-T (Tatsumi et al., 2019) and **(b)** MapCam (Rizk et al., 2018).

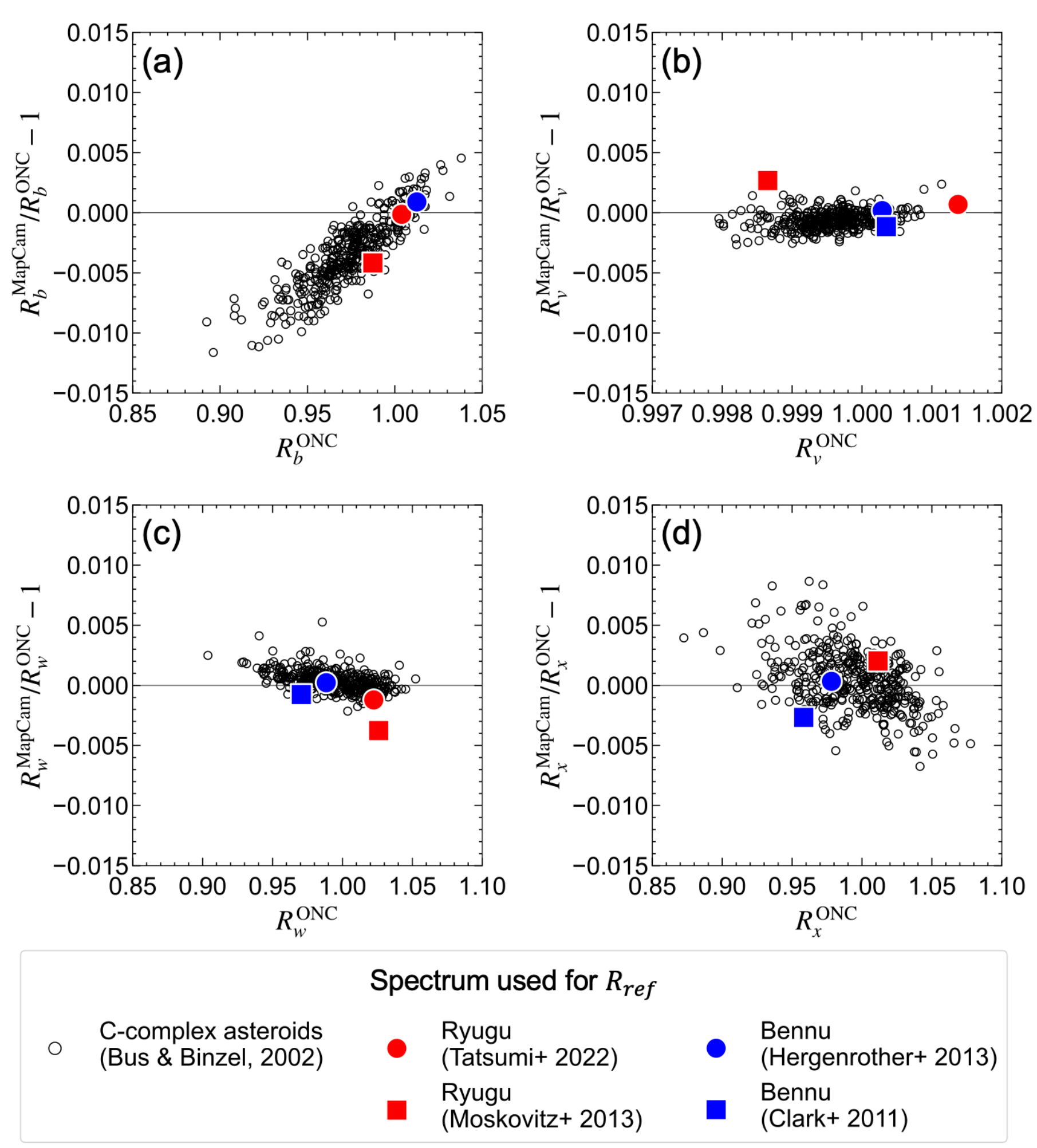

**Figure 3.** Difference between the simulated band spectrum of ONC-T ($\boldsymbol{R_n^{\mathrm{ONC}}}$) and MapCam ($\boldsymbol{R_n^{\mathrm{MapCam}}}$). Plots for each of the $n$ =$b$, $v$, $w$, and $x$ bands are shown in **(a)**−**(d)**.

**Table 2.** Values of $|\boldsymbol{R_n^{\mathrm{MapCam}}/R_n^{\mathrm{ONC}}-1}|$ (%), showing the differences in band spectra recorded by ONC-T and MapCam. $\boldsymbol{R_{ref}}$ shows the reference spectra used for evaluation.

| $R_{ref}$ | $b$ ($b'$) | $v$ | $w$ | $x$ |
|---|---|---|---|---|
| Ryugu (Tatsumi et al., 2022) | 0.01 | 0.07 | 0.12 | –† |
| Ryugu (Moskovitz et al., 2013) | 0.42 | 0.27 | 0.38 | 0.20 |
| Bennu (Hergenrother et al., 2013) | 0.09 | 0.02 | 0.02 | 0.03 |
| Bennu (Clark et al., 2011) | –† | 0.12 | 0.08 | 0.27 |
| C-complex asteroids<br>(RMSE among $N$ = 396 asteroid data;<br>Bus & Binzel, 2002a) | 0.42 | 0.10 | 0.09 | 0.24 |

† Exceeds the wavelength coverage of $R_{ref}$.

## 3. Bias between ONC-T and MapCam data before cross calibration

In this section, we identify the calibration parameters causing the bias between ONC-T and MapCam. The overall data reduction procedures are the same for both imagers and can be summarized as follows. First, we calibrate the raw image signals to "L1" images for MapCam (Golish et al., 2020a) and "L2b" images for ONC-T (Tatsumi et al., 2019). In these L1 and L2b images ($S_{obs,n}$), we remove bias/dark/read-out noise, apply flat-field correction, and divide by exposure time; $S_{obs,n}$ have units of digital counts per second (DN/s). Second, we radiometrically calibrate the images by dividing $S_{obs,n}$ by the radiometric calibration coefficients (RCC; Table 3), which we denote by $\mathrm{RCC}_n$. Third, we divide the radiometrically calibrated images by $J_n/\pi D^2$, where $J_n$ is the solar spectral irradiance at 1 au and $D$ is the sun-to-target distance in au, to convert them to radiance factor ($r_{obs,n}$; hereafter simply referred to as reflectance). These steps of calibration can be summarized as follows:

$$r_{obs,n} = \frac{S_{obs,n}}{\mathrm{RCC}_n}\frac{\pi D^2}{J_n}. \tag{2}$$

We label images of $r_{obs,n}$ as "iofL2" images for MapCam and "L2d" images for ONC-T, which are all accessible via the Planetary Data System.

ONC-T and MapCam independently derive the three calibration parameters in equation 2 (i.e., $D$, $J_n$, and $\mathrm{RCC}_n$). As we discuss below, inconsistencies in the definition and/or derivation of each parameter can lead to biased $r_{obs,n}$ between the two imagers. Our cross calibration is intended to resolve such inconsistencies and correct the imager-to-imager bias in $r_{obs,n}$.

- *Sun-to-target distance (D)*: For simplicity, MapCam uses the sun-to-spacecraft distance for $D$ in its calibration pipeline (Golish & Rizk, 2019), while ONC-T uses the sun-to-target distance. Since most images of Bennu were taken at low spacecraft altitudes, the effect of this approximation to the reflectance of Bennu is as small as <0.2% (Golish & Rizk, 2019), which is negligibly small in our cross calibration. Thus, we did not include the correction of $D$ in our cross-calibration method.

- *Solar spectral irradiance ($J_n$)*: The values of $J_n$ (Table 3) were calculated by weighted averages of a reference solar irradiance model ($J_{model}$; W/m$^2$/μm) as follows:

$$J_n = \frac{\int J_{model}(\lambda)\lambda\phi_n(\lambda)d\lambda}{\int \lambda\phi_n(\lambda)d\lambda}. \tag{3}$$

  ONC-T and MapCam use different models for $J_{model}$. ONC-T references the ASTM-e490 (ASTM, 2000) model ($J_{model}^{\mathrm{ONC}}$; Tatsumi et al., 2019) while MapCam references the Thuillier et al. (2004) model rescaled to have a solar constant of 1367 W/m$^2$ ($J_{model}^{\mathrm{MapCam}}$; DellaGiustina & Crombie, 2018). These two $J_{model}$ have local differences in depths of Fraunhofer lines (Fig. 4). For instance, $J_{model}^{\mathrm{MapCam}}$ exhibits a larger absorption of the Ca (II) line at 854 nm compared to $J_{model}^{\mathrm{ONC}}$. This inconsistency leads to a 2% imager-to-imager bias of $J_x$, leading to a 2% systematic difference in $r_{obs,x}$ of Ryugu and Bennu. Our cross-calibration method (section 4.1) compensates for such bias in $J_n$.

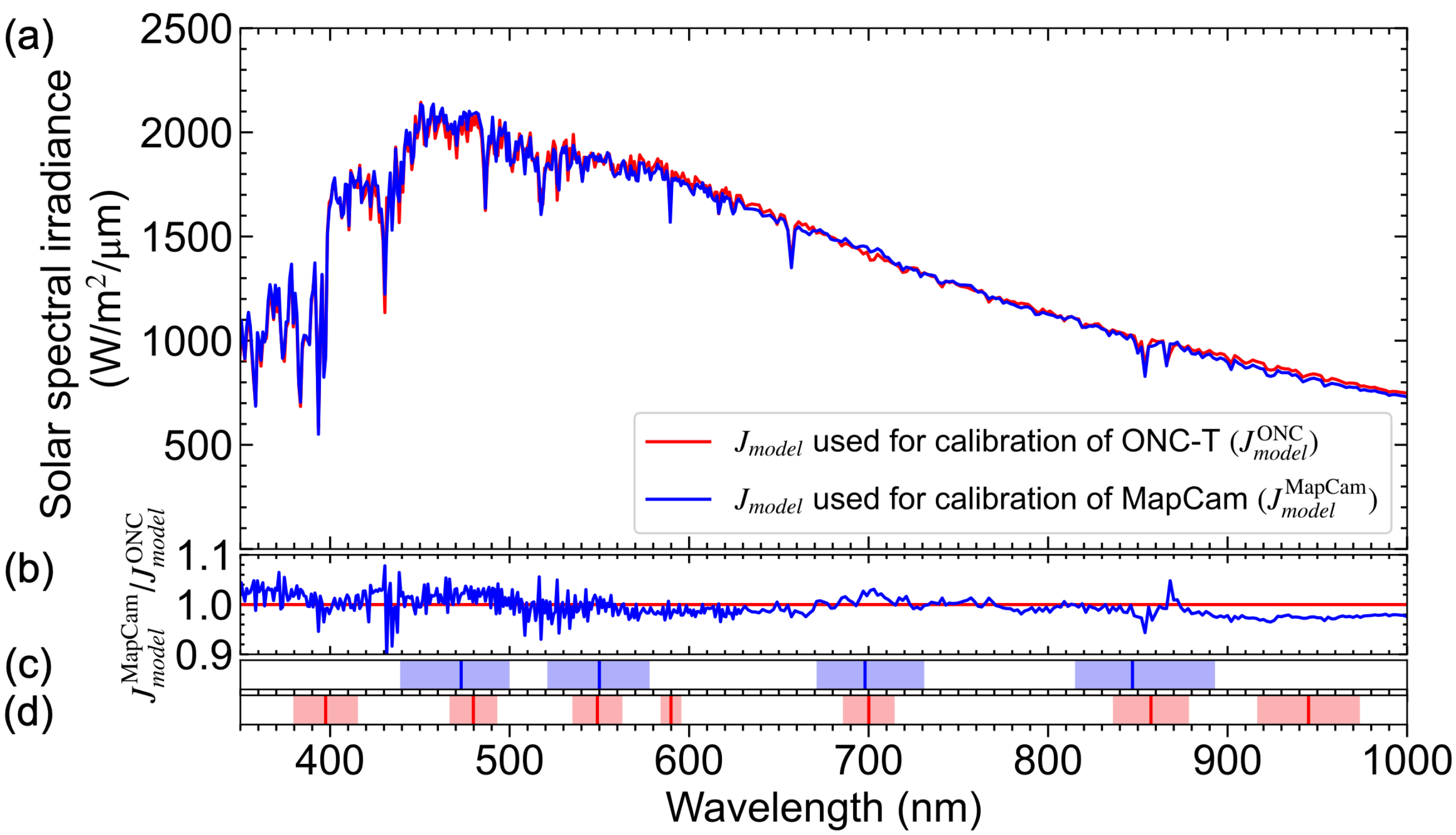

**Figure 4. (a)** Comparison of solar irradiance models used for calibration of ONC-T images ($J_{model}^{\mathrm{ONC}}$; red) and MapCam images ($J_{model}^{\mathrm{MapCam}}$; blue). **(b)** $J_{model}^{\mathrm{MapCam}}$ divided by $J_{model}^{\mathrm{ONC}}$. **(c)** The effective band centers (blue solid vertical lines) and the cut-on/cut-off wavelengths (hatches) of all four filters onboard MapCam (Golish et al., 2020a). **(d)** The effective band centers (red solid vertical lines) and effective band widths (hatches) of all seven filters onboard ONC-T (Tatsumi et al., 2019).

- *Radiometric calibration coefficient (*$\mathrm{RCC}_n$*)*: The values of $\mathrm{RCC}_n$ (Table 3) were derived from observation of light sources with known spectral irradiance. For both imagers, RCCs were updated during their flights by observations of natural light sources (e.g., stars), spectral irradiance of which is well documented.

  The RCC for ONC-T was determined based on the observations of eight standard stars (Tatsumi et al. 2019). The uncertainty in the RCC is a combination of two factors: (1) the precision of stellar observations by ONC-T, and (2) the accuracy in the measurements of absolute irradiance of standard stars by ground-based telescopes. We assess each of these error sources in the following.

(1) The evaluation of measurement precision is crucial because the observed intensities of point sources, such as stars, are known to vary significantly depending on their position within a detector. Precision becomes especially low when the detector has a low fill factor (i.e., the fraction of area sensitive to light). Nevertheless, the 100% fill factor of ONC-T, as illustrated in Fig. 1c, enables highly precise observations of stars.

We verified this expectation by calculating the precision among multiple images of the same standard star. We determined the position ($X$ and $Y$) of the star in each image by fitting a 2-D Gaussian. Subsequently, we calculated the stellar intensity by integrating the signal within a circular aperture with a radius of five pixels, after subtracting the background "sky" signal (mean within an annulus with an inner radius of 6 pixels and an outer radius of 8 pixels). Both the aperture and annulus were centered at the fitted Gaussian.

The observed intensities of the two brightest (V mag = 2–3) standard stars (Figs. 5) show that although each image observed the stars at various $(X, Y)$ positions within the detector due to spacecraft pointing jitter, their intensities remain stable within a 1σ precision of 0.5%. In addition, we observe no systematic variations with respect to $(X, Y)$. For the other six standard stars with V mag 3–5, the precision decreases due to increased shot noise and interference from cosmic rays, yet it remains below 2% (Fig. S1). The overall precision for the eight standard stars is 1.1% (1σ).

(2) Ground-based telescopes have measured the irradiances of standard stars with an accuracy of ~1.5−2.0% relative to Vega (α Lyrae) (Alekseeva et al., 1996). The absolute irradiance of Vega has been determined with an accuracy of ~1.0−1.5% by using terrestrial light sources with known output (e.g., black bodies) as a reference (Hayes, 1985). Consequently, the absolute irradiance of the standard stars has been determined with a combined uncertainty of 2–3%.

By propagating the errors introduced by (1) and (2), we determined the uncertainty in the RCC of ONC-T to be 3%. The reliability of this evaluation is supported by the consistency of RCCs derived from each of the eight standard stars, which demonstrated less than 3% variability (Tatsumi et al., 2019).

RCCs of MapCam were derived from observations of the Moon (Golish et al., 2020a). MapCam decided not to use stars because its relatively low fill factor of 76% (Fig.

1d) led to measurement precision of stars reaching tens of percent. Instead, MapCam utilized lunar data observed by the Robotic Lunar Observatory (ROLO) as the reference for calibration. The accuracy of ROLO data has been suggested to be 5–10% (Stone & Kieffer, 2004) or up to 8–13% (Velikodsky et al., 2011; Kieffer, 2022) based on its consistency with other lunar observatory/satellite data. Consequently, we estimate the uncertainty in the lunar-based RCC of MapCam to be 5–13%. Moon-based calibration typically has a larger error compared to stellar calibration due to the spatial variation and dependency on illumination and viewing geometries.

Consequently, bias between the RCCs of ONC-T and MapCam can reach 15%, because the two imagers used different calibration targets (i.e., stars vs the Moon) for their radiometric calibration. Although the band ratios of RCCs generally exhibit better accuracy than the RCCs themselves, we cannot rule out the possibility that band ratios may also be biased due to difference in calibration targets. Our cross-calibration method (section 4.1) corrects such bias in the RCCs.

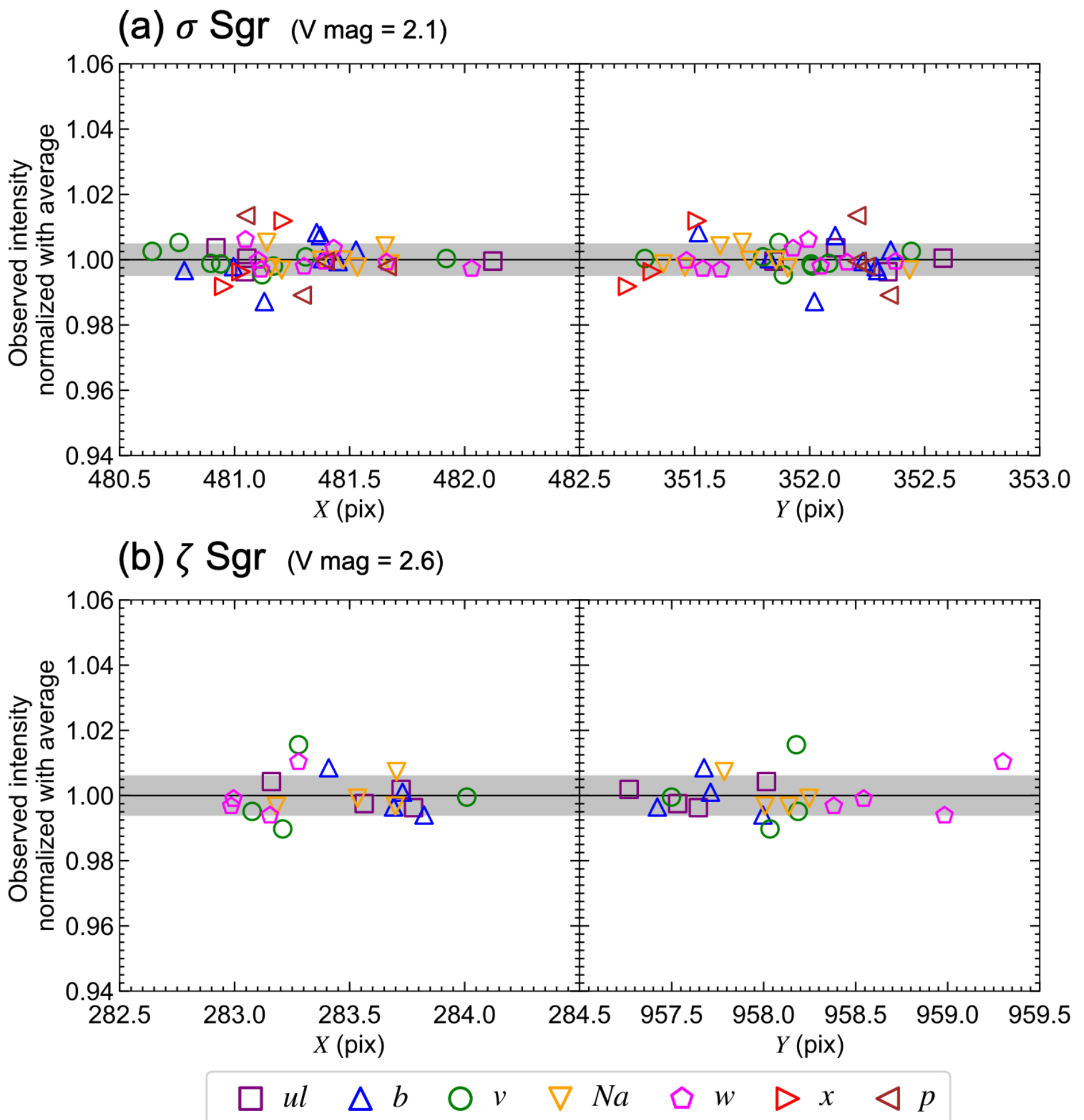

**Figure 5.** Precisions in the ONC-T observations of **(a)** σ Sgr and **(b)** ζ Sig, which are the two brightest standard stars used for the radiometric calibration of ONC-T. Each plot shows the observed intensity of the star in each image as functions of the star position in the **(left)** $X$ and **(right)** $Y$ coordinates of the detector. The intensities are normalized with the average for each band. The gray hatch shows the ±1σ range.

**Table 3.** Summary of the radiometric calibration coefficients ($\mathbf{RCC}_n$) and the solar spectral irradiance ($\boldsymbol{J}_n$) used in the calibration pipeline of each imager.

| | Band | $RCC_n$ (DN/s)/(W/m$^2$/μm/sr) † | $J_n$ (W/m$^2$/μm) † |
|---|---|---|---|
| ONC-T | *ul* | 439 | 1343.7 |
| | *b* | 969 | 1969.1 |
| | *v* | 1175 | 1859.7 |
| | *Na* | 547 | 1788.0 |
| | *w* | 1515 | 1414.4 |
| | *x* | 1500 | 985.8 |
| | *p* | 961 | 834.9 |
| MapCam | *b'* | 22,900 | 2003.2 |
| | *v* | 29,900 | 1837.8 |
| | *w* | 52,900 | 1426.9 |
| | *x* | 51,900 | 993.8 |

†Values were taken from Tatsumi et al. (2019) for ONC-T and Golish et al. (2020a) for MapCam.

## 4. Cross-calibration method

In section 4.1, we outline the concept and method for correcting the imager-to-imager bias discussed in section 3. In section 4.2, we discuss the sources of uncertainties associated with this method.

### 4.1 The bias correction factor ($\boldsymbol{F}_n$)

The goal of our cross calibration is to obtain the factor ($F_n$) that corrects the bias between the reflectance of Bennu observed by MapCam and those of Ryugu observed by ONC-T for each of the $n = b$, *v*, *w*, and *x* band. The uncertainties of $F_n$ constrain the *relative* multiplicative difference between the reflectance and spectral shape of Ryugu and Bennu.

In contrast, $F_n$ does not constrain the *absolute* values of the cross-calibrated spectra. We decided to rely on ONC-T data for the absolute calibration due to its higher radiometric accuracy (section 3). We further validate this approach in section 7.

Thus, we define $F_n$ as the factor correcting for the offset of MapCam to ONC-T. The cross-calibrated reflectance of MapCam ($r'^{\mathrm{MapCam}}_{obs,n}$) and ONC-T ($r'^{\mathrm{ONC}}_{obs,n}$) can be obtained by multiplying the pre-cross-calibrated reflectance of MapCam data ($r^{\mathrm{MapCam}}_{obs,n}$) by $F_n$ while retaining those of ONC-T data ($r^{\mathrm{ONC}}_{obs,n}$):

$$r'^{\mathrm{MapCam}}_{obs,n} = F_n \cdot r^{\mathrm{MapCam}}_{obs,n}; \quad r'^{\mathrm{ONC}}_{obs,n} = r^{\mathrm{ONC}}_{obs,n}. \tag{4}$$

As shown in section 3, $F_n$ needs to include two components $f_{J_n}$ and $f_{\mathrm{RCC}_n}$, which respectively corrects for the difference in solar irradiance models and targets for radiometric calibration:

$$F_n = f_{J_n} \cdot f_{\mathrm{RCC}_n}. \tag{5}$$

These factors $F$, $f_{J_n}$, and $f_{\mathrm{RCC}_n}$ normalized at those at *v* band are denoted as $\widehat{F_n}$, $\widehat{f_{J_n}}$, and $\widehat{f_{\mathrm{RCC}_n}}$, respectively and give the correction factors for band ratios. The methods for obtaining $f_{J_n}$ and $f_{\mathrm{RCC}_n}$ are described in the following sections 4.1.1 and 4.1.2, respectively.

#### 4.1.1 The solar irradiance correction factor ($f_{J_n}$)

We recalculated the solar spectral irradiance for MapCam by using $J^{\mathrm{ONC}}_{model}$ instead of $J^{\mathrm{MapCam}}_{model}$ in equation 3. Using the original ($J^{\mathrm{MapCam}}_n$) and recalculated ($J'^{\mathrm{MapCam}}_n$) solar spectral irradiance, we obtain $f_{J_n}$ using the following equation:

$$f_{J_n} = \left(J'^{\mathrm{MapCam}}_n / J^{\mathrm{MapCam}}_n\right)^{-1}. \tag{6}$$

Here, the ratio is to the minus one because reflectance is proportional to $J_n^{-1}$ as shown in equation 2.

It is noted that although solar irradiance varies over time, daily data by the SORCE satellite (Harder, 2020) show that such time variation is negligible; the peak-to-peak variation is 0.2% for all *b*, *v*, *w*, and *x* bands within the time range of observation (Fig. S2).

#### 4.1.2 The RCC correction factor ($f_{\mathrm{RCC}_n}$)

We calculated $f_{\mathrm{RCC}_n}$ based on observations of a common calibration target. Ideally, a calibration target should exhibit consistent irradiance across different temporal, spatial, and illumination/viewing geometries. In the event of any variations, it becomes imperative to employ corrective models to mitigate these effects. In contrast, biases intrinsic to the reference spectra of calibration targets are not critical because the main goal of our cross calibration is to accurately derive the relative signal intensity between the imagers.

Although irradiances of standard stars are generally constant, these are not the best targets in our case due to the low fill factor of MapCam (section 3). Other distant planets, such as Mars,

Jupiter, and Saturn, are unsuitable for the same reason.

We used the Moon as the target for cross calibration. Both imagers took spatially resolved multi-band images of the Moon during the Earth swing-by operations before arriving at the asteroids. The conditions of lunar observations are summarized in Table 4. The Moon is a suitable target because it is well extended in the field-of-view (FOV) and the lunar radiance is stable over time ($10^{-5}$%/yr; Kieffer, 1997). However, the two imagers observed almost opposite areas of the Moon (Fig. 6) with different illumination and viewing geometries (i.e., incidence, emission, and phase angles: $i$, $e$, and $\alpha$). Since the radiance of the Moon is spatially inhomogeneous due to variations in albedo and dependent on ($i$, $e$, $\alpha$), the lunar images taken by the two imagers need to be compared through the same lunar photometry model, which provides spatially- and disk-resolved reference spectra of the Moon. It is noted that ONC-T also observed the Moon five years after its initial lunar campaign during the sample capsule separation in 2020. Since these images are affected by a 10% decrease in sensitivity caused by the touchdown operations (Yamada et al., 2023), we only used them for checking the robustness of our band ratio calibration.

**Table 4.** Summary of lunar images used in our analysis and their observation geometries.

| Imager | Obs. Date | Number of multi-band images[†] | Solar phase angle at the Moon center (°) | Moon–sun distance (AU) | Moon–spacecraft distance (km) | Moon diameter in image (pix) | Sub-spacecraft latitude and longitude | Sub-solar latitude and longitude |
|---|---|---|---|---|---|---|---|---|
| ONC-T | Dec 5, 2015 | 3 (7 bands each)[§] | 59.3 | 0.984 | 764,000–774,000 | 42 | 56°S, 263°E | 2°N, 248°E |
| | Dec 6, 2020 | 3 (7 bands each) | 49.4–50.2 | 0.986 | 547,000–573,000 | 56–59 | 7°N, 335°E | 0°N, 286°E |
| MapCam | Sep 25, 2017 | 12 (4 bands each) | 41.6–42.3 | 1.001 | 1,170,000–1,245,000 | 41–44 | 33°N, 96°E | 1°N, 125°E |

† See Fig. 9 for the location and size of the lunar image in the FOV.

§ For one multi-band image set, the Moon was imaged only 50 pix away from the corner of the FOV. This image set was excluded from our analysis due to the large errors in flat fielding (section 4.2).

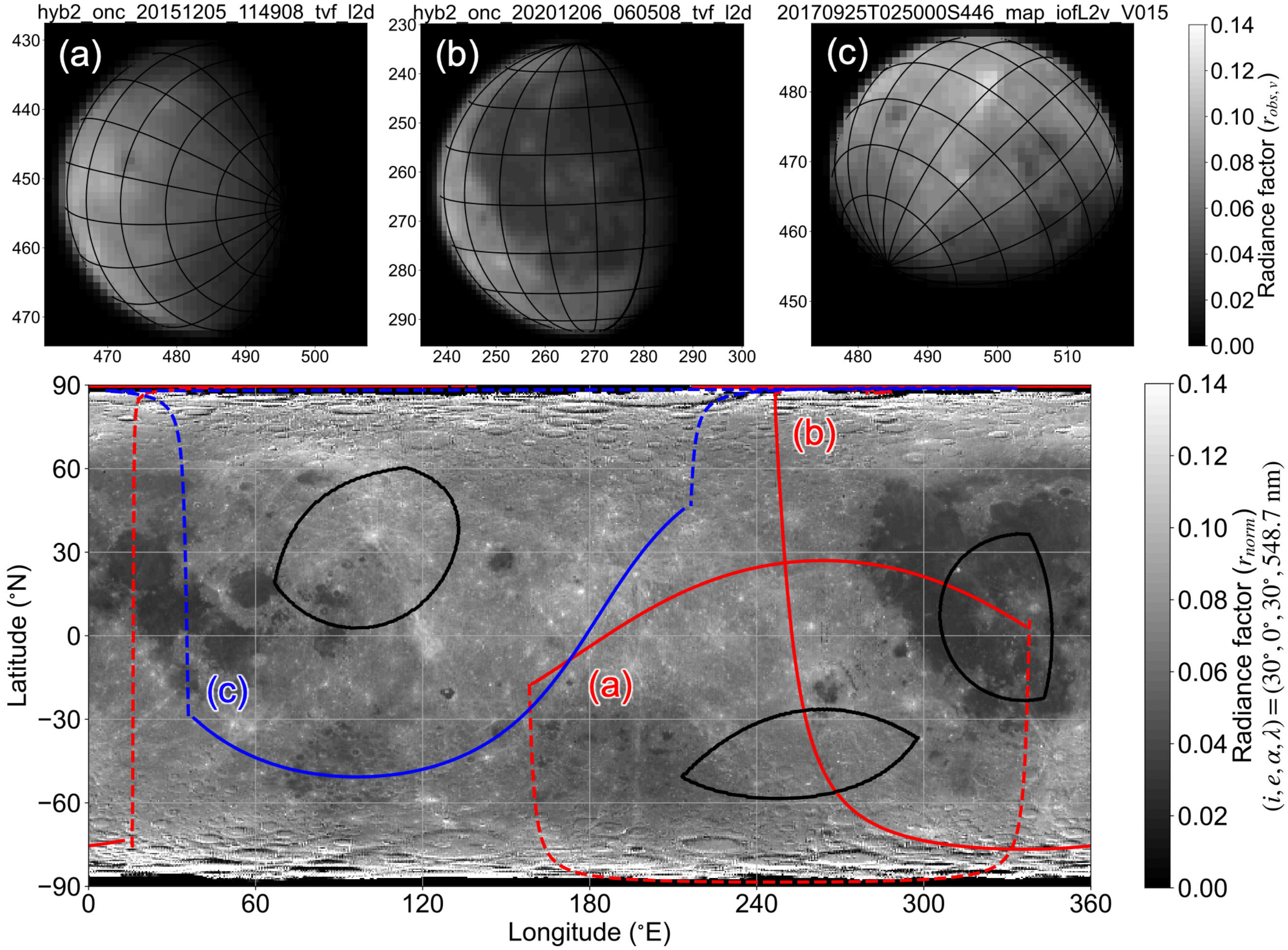

**Figure 6.** Lunar images taken by ONC-T and MapCam and their imaged area on the lunar map. Top three images show the observed lunar images ($\boldsymbol{r_{obs,v}}$) taken by **(a)** ONC-T during the cruise enroute to Ryugu on Dec 5, 2015, **(b)** ONC-T during the returning cruise on Dec 6, 2020, and **(c)** MapCam during the cruise enroute to Bennu on Sep 25, 2017. Solid curves show contours of latitudes and longitudes. Their imaged regions are plotted on the photometrically-normalized SP-model lunar reflectance map **(bottom)**, indicating that ONC-T and MapCam observed opposite areas of the Moon. The red and blue solid curves depict the limb of the Moon for each of the observed images shown in the top row, while the dashed curves represent the terminators. The black solid curves show regions observed with incidence angles of <60º and emission angles of <30º.

We simulated each image of the Moon observed by ONC-T and MapCam using two lunar photometry models, following the method developed in Kouyama et al. (2016). Since ONC-T and MapCam mostly observed the far-side of the Moon (Fig. 6), simulation using photometric models developed from spacecraft data have advantages over those from ground-based telescope observations owing to its larger compatibility in the observed region and geometry. We incorporated two photometric models into this study to cover the entire wavelength range (i.e., 0.48–0.85 μm) and to evaluate their consistency. These two models were the Lunar Reconnaissance Orbiter/Wide Angle Camera model (Sato et al., 2014; here after WAC model) and the Kaguya/Spectral Profiler model (Yokota et al., 2011; Kouyama et al., 2016; here after SP model). The WAC model provides disk-resolved spectrophotometry for seven spectral bands at 321, 360, 415, 566, 604, 643, and 689 nm, which could be used to simulate the *b* and *v*-band images. The SP model provides disk-resolved spectrophotometry from 520 to 2600 nm with a spectral resolution of ~6 nm and complies with the *v*, *w*, and *x* band. For the WAC model, Sato et al. (2014) fitted the photometric data of each 1°E × 1°N resolution mesh on the lunar surface with photometric functions developed by Hapke et al. (2012). In contrast, for the SP model, Yokota et al. (2011) classified the lunar surface into three albedo groups (high, medium, and low albedo) and fitted the photometric data for each group with photometric functions developed by McEwen (1991, 1996). We summarized the simulation procedure graphically in Fig. 7 and described it in appendix A. The simulated images ($r_{sim,n}$) are compared with the observed images ($r_{obs,n}$) in Fig. 8.

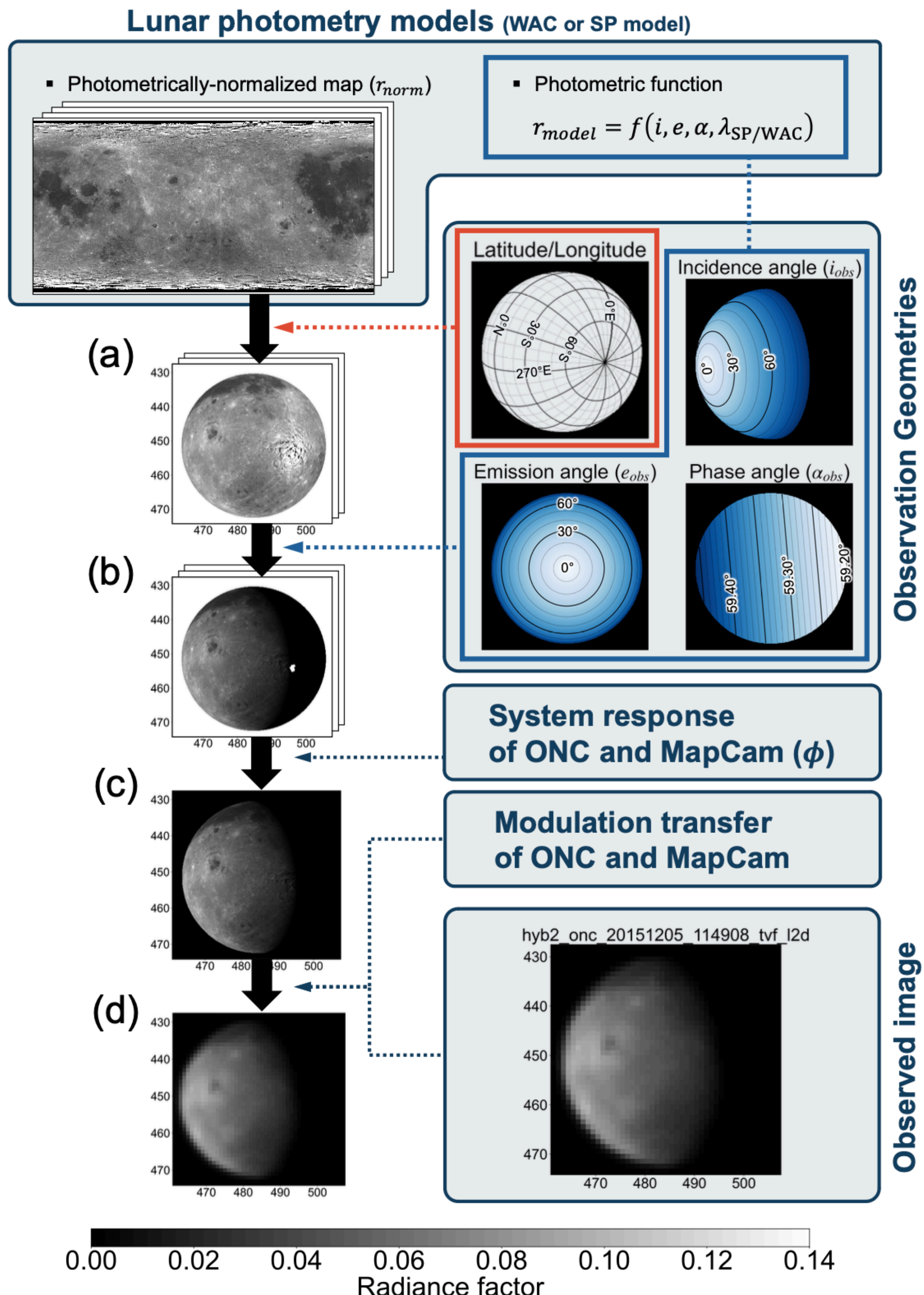

**Figure 7.** Procedure for simulating the observed image of the Moon using the lunar photometry models. **(a)** Project a map of photometrically-normalized reflectance to the observed image plane

using geometric information. **(b)** Photometrically correct the reflectance to the observation condition. **(c)** Use the spectral response function of each filter on ONC-T and MapCam to calculate the in-band reflectance. Fill "sky" background and regions with data deficiency with zeros. **(d)** Down-sample the image to the observed resolution and convolve it with the point spread functions of ONC-T and MapCam. Iterate on steps (c) to (d) varying image shift and rotation to obtain the best co-registration accuracy with the observed image.

We obtained $f_{\mathrm{RCC}_n}$ by averaging $r_{sim,n}/r_{obs,n}$ (Fig. 8c) over the lunar disk ($\overline{r_{sim,n}/r_{obs,n}}$) and calculating the imager-to-imager ratio:

$$f_{\mathrm{RCC}_n} = \left[\overline{r_{sim,n}/r_{obs,n}}\right]^{\mathrm{MapCam}} / \left[\overline{r_{sim,n}/r_{obs,n}}\right]^{\mathrm{ONC}}. \quad (7)$$

We used $J_{model}^{\mathrm{ONC}}$ for $J_{model}$ and the sun-to-Moon distance for $D$ in deriving the $r_{obs,n}$ in equation 7. We derived $\widehat{f_{\mathrm{RCC}_n}}$ by normalizing $\overline{r_{sim,n}/r_{obs,n}}$ by its value at *v* band for each multi-band image set.

We only averaged pixels with high photometric correction accuracy and photon-to-signal linearity, based on the following criteria. The ratio image (Fig. 8c) shows local anomalies at regions observed with high incidence or emission angles. This is because reflectance observed with emission angles larger than ~30º and incidence angles larger than ~60º cannot be accurately simulated using the WAC and SP models (Sato et al., 2014; Kouyama et al., 2016). Thus, we only averaged pixels with incidence angles <60º and emission angles <30º. In addition, only pixels with DN values ranging from 1,000 to 13,000 were used for MapCam images to exclude pixels with >0.5% non-linearity (Golish et al., 2020a). For ONC-T images, all pixels had DN values <3200, where the linearity is better than 0.6% (Tatsumi et al., 2019). For each image, a total of >150 pix met these criteria and the pixel-to-pixel correlation between $r_{sim,n}$ and $r_{obs,n}$ exceeded 0.98 (Fig. 8d).

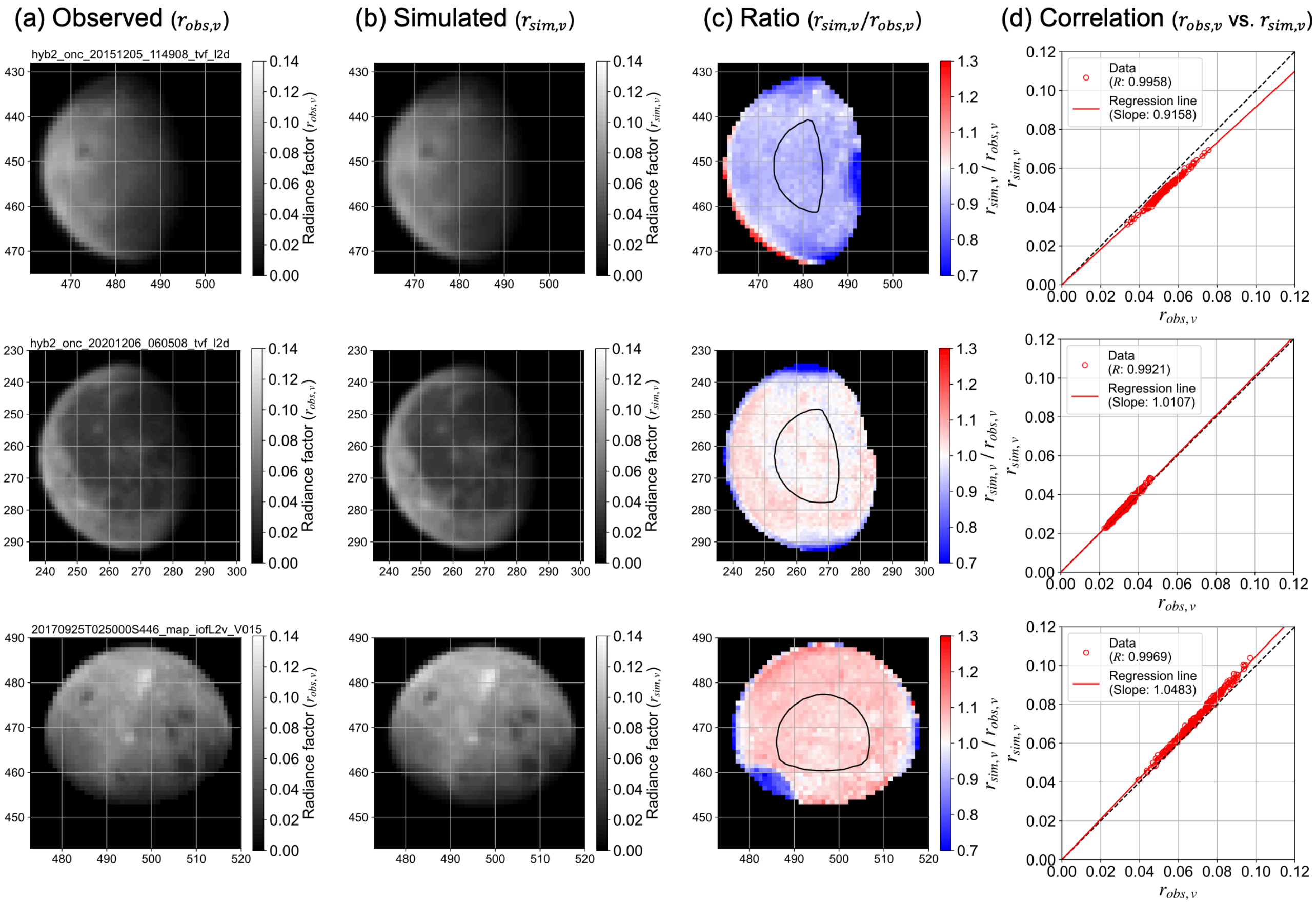

**Figure 8.** Comparison between **(a)** the observed image ($\boldsymbol{r_{obs,n}}$) and **(b)** image simulated using the WAC model ($\boldsymbol{r_{sim,n}}$) for the **(top row)** ONC-T observations after launch (Dec 5, 2015), **(middle row)** ONC-T observations during the returning cruise (Dec 6, 2020), and **(bottom row)** MapCam observations (Sep 25, 2017). The ratio image ($\boldsymbol{r_{sim,n}/r_{obs,n}}$) is shown in **(c)**. Pixels within the dark curve (incidence angle <60º, emission angles <30º, and latitudes <70ºS – 70N) were averaged. Pixel-by-pixel correlation between $\boldsymbol{r_{obs,n}}$ and $\boldsymbol{r_{sim,n}}$ within the dark curve is shown in **(d)**. The same plots with $\boldsymbol{r_{sim,n}}$ based on the SP model are shown in Fig. S6.

### 4.2 Uncertainties in the bias correction factor ($\boldsymbol{F_n}$)

We evaluated the uncertainties in $F_n$ (and $\widehat{F_n}$) caused by the following three error sources.

A) Errors in $f_{\mathrm{RCC}_n}$ due to modelling errors in $r_{sim,n}$

B) Errors in $f_{\mathrm{RCC}_n}$ due to flat-fielding errors and noise in $r_{obs,n}$

C) Difference in spectral responsivities between the two imagers limiting their direct comparison

We label the errors accounting for each of these sources as $\sigma_A$, $\sigma_B$, and $\sigma_C$, which we respectively evaluate by the following methods A–C. We calculate the overall uncertainty in $F_n$ by their root sum of squares.

(A) Errors in photometric correction using the WAC and SP models (equation 10 in appendix A) lead to inaccurate $f_{\mathrm{RCC}_n}$. We evaluated this error ($\sigma_A$) to be 0.9% by comparing $f_{\mathrm{RCC}_n}$ derived from the WAC and SP models. This approach is valid because the two models employ different photometric functions. The value of $\sigma_A$ could only be evaluated at *v* band where the wavelength coverages of WAC and SP models overlap. Nevertheless, the value is representative for other bands because the same photometric functions are used.

We note that the systematic model-to-model difference in the phase function is not included in our evaluation of $\sigma_A$. Instead, we applied a correction function ($q(\alpha)$ in appendix A) to the phase function of the SP model to align it with that of the WAC model, as the phase function of the WAC model is more accurate (appendix A). We estimated $\sigma_A$ for $\widehat{F_n}$ to be 0.7% from the wavelength dependency of $q(\alpha)$ (Fig. S4b).

(B) The uncertainty in $f_{\mathrm{RCC}_\mathrm{n}}$ caused by image noise is minimized in our cross calibration by averaging over >150 pixels within the lunar disk (section 4.1.2). Similarly, the uncertainty caused by errors in flat fielding is minimized by using different image sets that observed the lunar disk at different positions within the FOV (Fig. 9). Nevertheless, we evaluate these uncertainties ($\sigma_B$) by the standard deviation of $f_{\mathrm{RCC}_\mathrm{n}}$ derived from different image sets of the Moon (Table 4), finding that these uncertainties are 1%. We did exclude one ONC-T image set from our analysis where the Moon was imaged at the corner of the FOV because the flat-fielding accuracy is significantly low in the peripheral regions; ONC-T has a large vignetting of -60% (Kameda et al., 2017) at the corner of its FOV. Thus, the estimation of $\sigma_B$ is valid within the central ~600 pixels for ONC-T and central ~400 pixels for MapCam at which the lunar disk was imaged (Fig. 9).

Similarly, we evaluated $\sigma_B$ for $\hat{F}$ to be 0.5% based on the standard deviation of $\widehat{f_{\mathrm{RCC}_n}}$ derived from different image sets. The $\sigma_B$ becomes smaller for $\hat{F}$ because any band-to-band correlated errors (e.g., most of the errors in flat-fielding) are cancelled out in $\widehat{f_{\mathrm{RCC}_n}}$.

We note that although the effect of time-invariant noise, mainly stray light, are not included in our estimation of $\sigma_B$, they are negligible for lunar images. We evaluated the effects of stray lights to be smaller than 0.3% based on the ratio of the observed background intensity to the brightness of the lunar disk.

(C) Even after all imager-to-imager bias in the calibration parameters are corrected, slight difference in the spectral responsivities limits the direct comparison of ONC-T and MapCam data as discussed in section 2. We account for this limitation by including an error ($\sigma_C$) in $F_n$. We conservatively estimated $\sigma_C$ to be 0.3% by taking the largest value in Table 2. We evaluated the value of $\sigma_C$ for $\widehat{F_n}$ to be 0.4% based on error propagation.

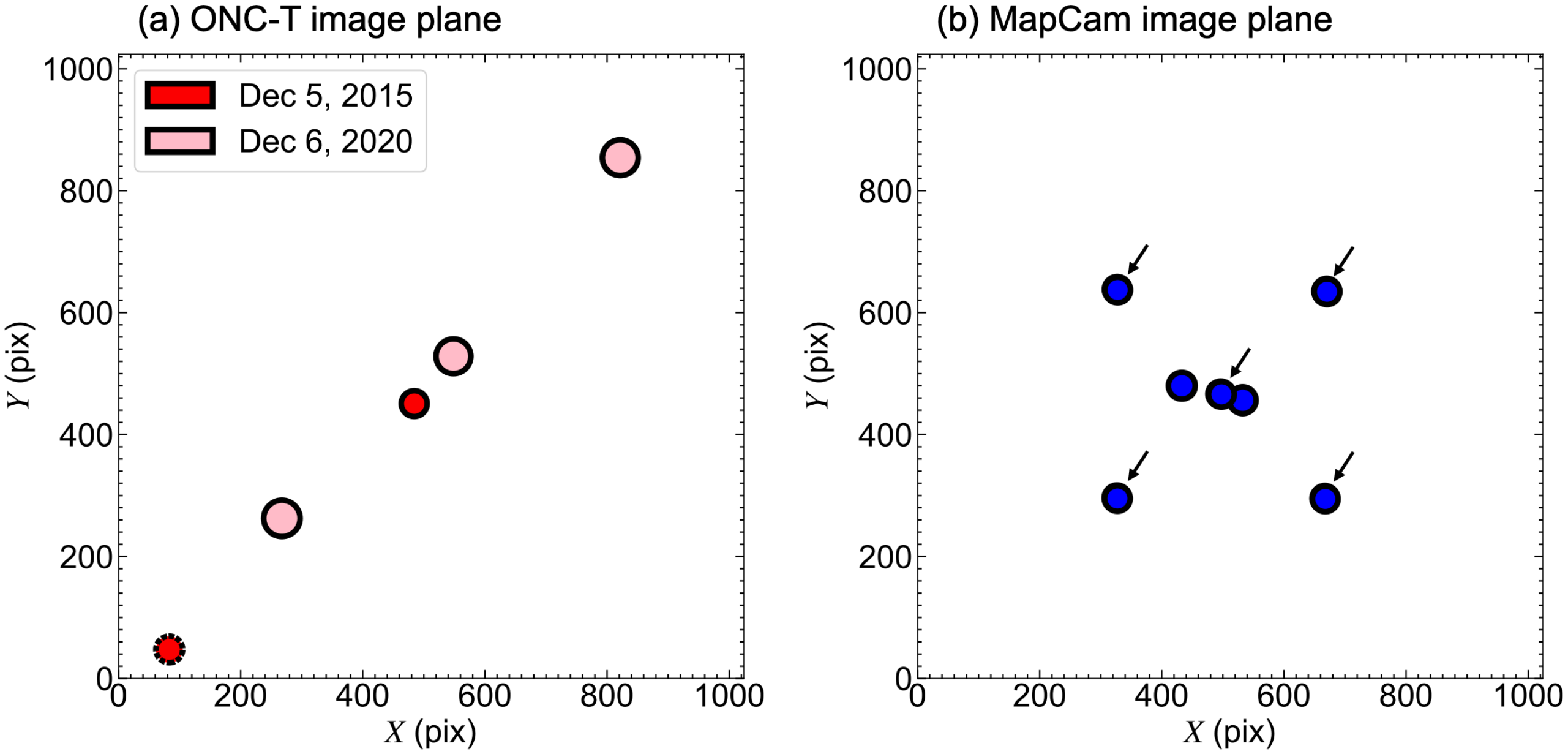

**Figure 9.** Location and size of the lunar image in the FOV for all observations by **(a)** ONC-T and **(b)** MapCam. The position and size of each circle represent those of the lunar disk in each multi-band image set. One ONC-T image set plotted with dashed outlines was excluded from our analysis due to off-centering (see section 4.2). Arrows show two overlapping plots.

## 5. Results: systematic calibration difference between ONC-T and MapCam

We first examine the simulation-to-observation ratio of lunar images ($\overline{r_{sim,n}/r_{obs,n}}$). The deviations of $\overline{r_{sim,n}/r_{obs,n}}$ from unity are <5% for all bands of MapCam while those are 6–15% for ONC-T (Fig. 10a). The band ratios of $\overline{r_{sim,n}/r_{obs,n}}$ derived from the two observation campaigns by ONC-T are consistent within 2% (Fig. 10b), showing the robustness of our result; most of the 2% difference can be explained by the non-uniform degradation at different bands (Kouyama et al., 2021).

The MapCam observations ($r_{obs,n}$) are closer to the simulated lunar data ($r_{sim,n}$) because they share the same systematic error, and it does not necessarily indicate that the radiometric calibration accuracy of MapCam is higher. As discussed in section 3, the original calibration of MapCam is based on the photometric model developed from the ROLO data. Similarly, the WAC and SP data, from which our $r_{sim,n}$ are derived, were calibrated using the ROLO data (see appendix A). Consequently, any systematic errors inherent to the ROLO data are shared between MapCam data and the simulated lunar data. The reason why MapCam data does not perfectly agree with the simulated lunar data can be attributed to the fact that the original calibration of MapCam only used the nearside of the lunar surface covered by ROLO, whereas our analysis uses the entire illuminated surface. Comparison between ROLO and WAC data by Mahanti et al. (2016) shows that even though these two data are consistent in their average lunar radiance, their representation of the spatial brightness distribution have a discrepancy of 4–6%.

In contrast, ONC-T data is not influenced by any systematic errors in the ROLO data because the calibration of ONC-T is based on standard stars. Fig. 10a indicates a 6−15% systematic difference between the stellar data used for the original radiometric calibration of ONC-T and the lunar data.

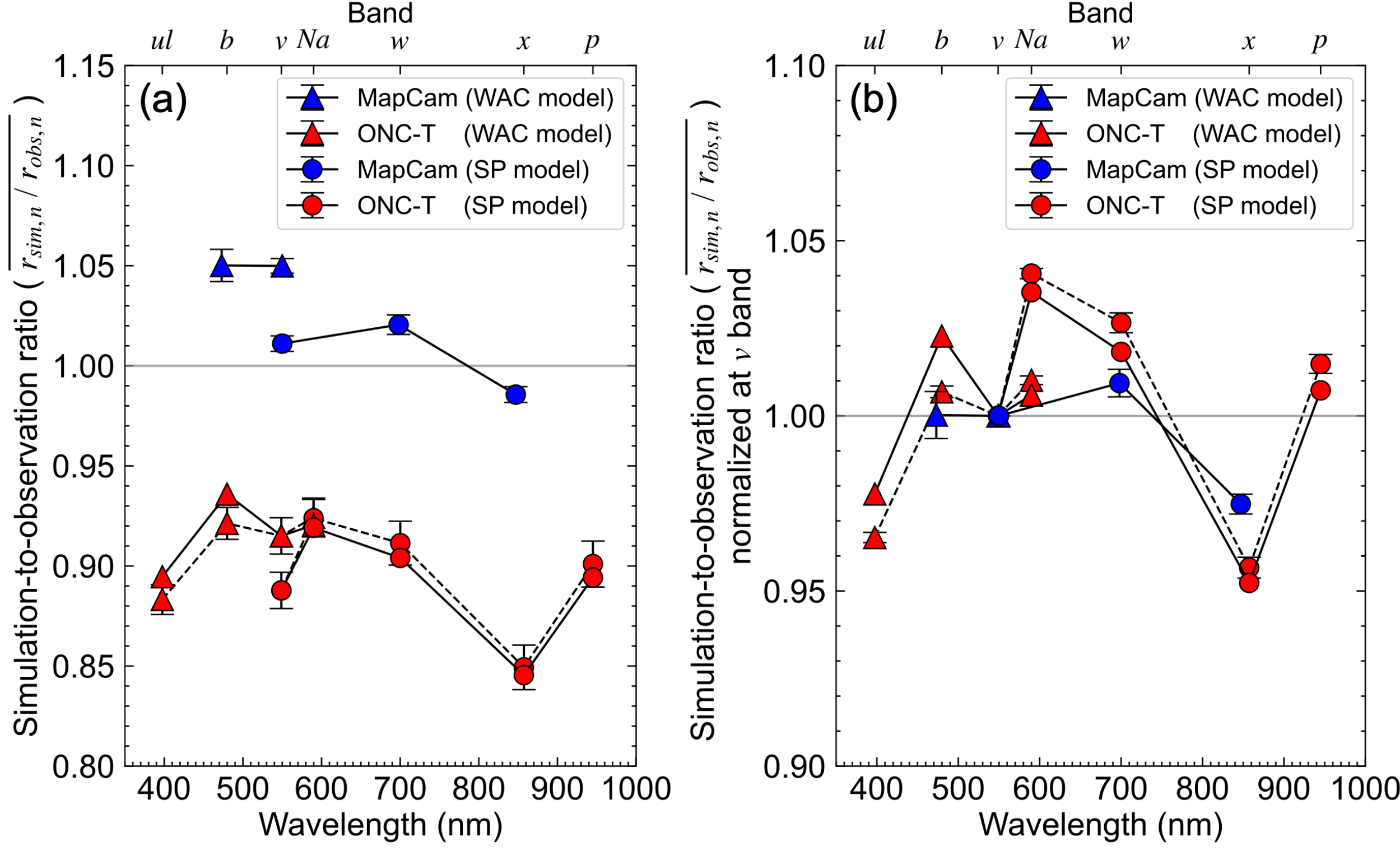

**Figure 10.** Simulation-to-observation ratio of the lunar reflectance image ($\overline{r_{sim,n}/r_{obs,n}}$). The values of $\overline{r_{sim,n}/r_{obs,n}}$ are shown in **(a)** and those normalized at *v* band are shown in **(b)**. Error bars show the precision evaluated from different images of $r_{obs,n}$: $\sigma_B$ in section 4.2. For ONC-T, the results obtained from observations after launch (Dec 5, 2015) are shown in solid lines whereas those from observations during the returning cruise (Dec 6, 2020) are shown in dashed lines. We offset the $\overline{r_{sim,n}/r_{obs,n}}$ derived from the images taken during the return cruise to match the value after launch at *v* band to correct for the 10% degradation caused by the touchdown operations (Kouyama et al., 2021).

The values of $f_{\mathrm{RCC}_n}$ (i.e., the imager-to-imager ratio of $\overline{r_{sim,n}/r_{obs,n}}$ ; red lines in Fig. 11) more clearly shows that the ROLO data used for the calibration of MapCam is systematically lower than the standard star data used for the calibration of ONC-T by 13% for all *b*–*x* bands. Although the exact cause of this discrepancy remains unidentified, our result is consistent with Velikodsky et al. (2011) and Kieffer (2022), which point out that lunar radiance observed by ROLO is systematically lower by 8–13% compared to that observed in other facilities. Therefore, the 13% discrepancy between the ROLO and stellar data, which led to the significant bias between ONC-T and MapCam, might entirely result from a systematic error specific to the ROLO data. Our results underscore the need to reconcile and improve the reference data of the Moon for more accurate instrument calibration in future space missions.

We obtain the overall bias correction factor $F_n$ (black lines in Fig. 11) by multiplying ${f_J}_n$ (blue lines in Fig. 11) to $f_{\mathrm{RCC}_n}$ (red lines in Fig. 11). The values of $f_{\mathrm{RCC}_n}$, ${f_J}_n$, and $F_n$ respectively show how much correction is needed to resolve the imager-to-imager bias in RCC, solar irradiance, and both RCC and solar irradiance. The values of $F_n$ show that the pre-cross-calibrated data had a significant imager-to-imager bias of 13.2 ± 1.5% in reflectance at *v* band (Fig. 11a). This bias is twice as large as the difference in reflectance between Ryugu and Bennu evaluated from the pre-cross-calibrated data (DellaGiustina et al., 2020). Thus, correcting this bias with $F_n$ is crucial for an accurate comparison of reflectance.

In contrast, the values of $\widehat{F_n}$ (Fig. 11b) indicate that imager-to-imager bias in band ratio was as small as <1.5%. The uncertainties in $\widehat{F_n}$ were 1.1% for the *b*/*v* band ratio and 0.9% for the *w*/*v* and *x*/*v* band ratios, equivalent to an uncertainty of ±0.032 μm$^{-1}$ in the *b*-to-*x*-band spectral slope (see Yumoto et al., 2024 for definition). For the *b*/*v* and *w*/*v* band ratios, we detect no significant imager-to-imager bias exceeding the uncertainties of our cross calibration. Although we detect a bias larger than the uncertainty for the *x*/*v* band ratio, it only has a minor effect because the bias is smaller than 1/5 the global difference between Ryugu and Bennu.

Our results conclude that absolute reflectance measured by ONC-T and MapCam had large imager-to-imager bias mainly due to difference in their radiometric calibration targets, but bias in the band ratio was nearly negligible (<1.5%).

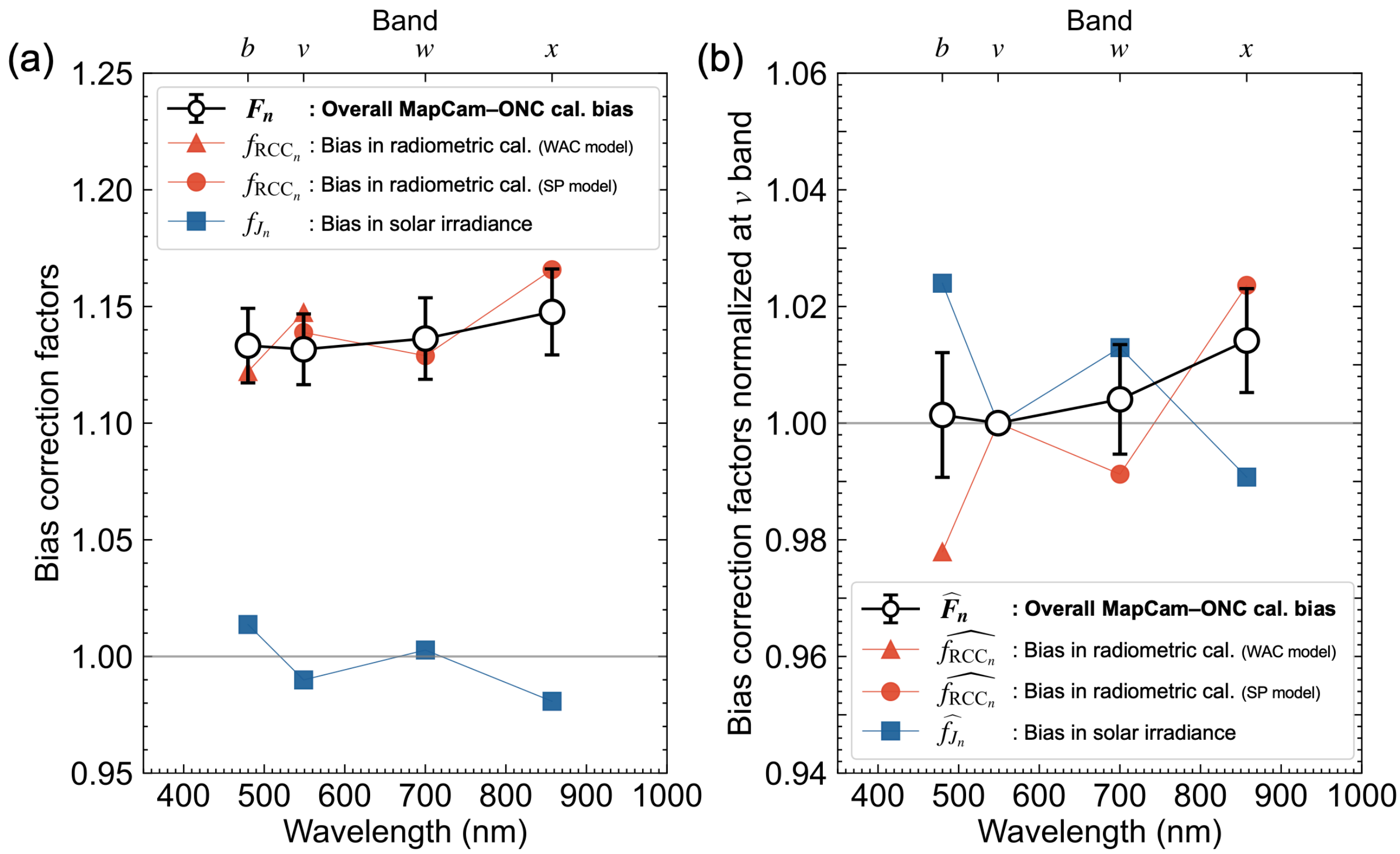

**Figure 11.** Correction factors compensating for the imager-to-imager bias. The factor $\boldsymbol{f_{J_n}}$ corrects the bias caused by the difference in solar irradiance models (section 4.1.1) and $\boldsymbol{f_{\mathrm{RCC}_n}}$ corrects the difference in radiometric calibration targets (section 4.1.2). The overall bias correction factor ($\boldsymbol{F_n}$) is $\boldsymbol{f_{J_n} \cdot f_{\mathrm{RCC}_n}}$. The error bars show $\sqrt{\boldsymbol{\sigma_A^2 + \sigma_B^2 + \sigma_c^2}}$ discussed in section 4.2. Values of these factors are shown in **(a)** and those normalized at *v* band are shown in **(b)** (refer to Table 5 for the data).

## 6. Correction of the imager-to-imager bias for future comparative studies of Ryugu and Bennu

Future comparative studies of Ryugu and Bennu can correct the bias of MapCam data relative to ONC-T data and obtain the cross-calibrated reflectance by multiplying the pre-cross-calibrated reflectance of Bennu by the values of $F_n$ summarized in Table 5. No correction is needed for the reflectance of Ryugu as shown in equation 4. The pre-cross-calibrated data are publicly available via the Planetary Data System: "iofL2" images (urn:nasa:pds:orex.ocams:data_calibrated; version 13.0; Rizk et al., 2019) for the data of Bennu and "L2d" images (urn:jaxa:darts:hyb2_onc:data_iof; version 1.0; Sugita et al., 2022) for the data of Ryugu. The <2% uncertainties in $F_n$ (Table 5) constrain the *offset* between the cross-calibrated reflectances of Ryugu and Bennu while their *absolute* reflectances are constrained by the star-based radiometric calibration accuracy of ONC-T, which is 3% (section 3 and validated in section 7).

Similarly, we can obtain the cross-calibrated band ratios of Ryugu and Bennu by multiplying the pre-cross-calibrated band ratios of Bennu by $\widehat{F_n}$ (Table 5). We note that applying $\widehat{F_n}$ to the pre-cross-calibrated band ratios is equivalent to applying $F_n$ to the pre-cross-calibrated reflectance and then normalizing it at *v* band, but either way, the uncertainties in $\widehat{F_n}$ (Table 5) show that we can compare the band ratios of Ryugu and Bennu with accuracies of 0.9−1.1% after applying our cross calibration.

Since $F_n$ is specific to the instruments, it can be also applied to the data of asteroids encountered during the extended missions of Hayabusa2 and OSIRIS-REx. Hayabusa2 will fly by the asteroid (98943) 2001 $CC_{21}$ in 2026 and arrive at 1998 $KY_{26}$ in 2031 (Hirabayashi et al., 2021). The extended mission for OSIRIS-REx, called OSIRIS-APEX, will arrive at asteroid (99942) Apophis in 2029 (DellaGiustina et al., 2022). However, the ~10% change in instrument response caused by the sample collection performed during the final stages of Ryugu and Bennu survey (Kouyama et al., 2021; Lauretta et al., 2022) needs additional correction. This also caused the instrument response to vary over months and years after the sample collection (Kouyama et al., 2021; Yamada et al., 2023). Thus, at this moment, we cannot determine the exact correction factor needed for comparing the ONC-T and MapCam data collected during the extended mission phase. The sensitivities of ONC-T and MapCam will be continuously monitored by stellar and lunar observation campaigns throughout the extended missions.

**Table 5.** Values and uncertainties in the bias correction factor ($\boldsymbol{F_n}$) for each of the $\boldsymbol{n} =$ *b*, *v*, *w*, and *x* bands shared by ONC-T and MapCam. The cross-calibrated reflectance of Ryugu and Bennu can be obtained by multiplying the pre-cross-calibrated reflectance of Bennu by these values while retaining the reflectance of Ryugu.

Bias correction factor ($F_n$)

| Bands ($n$) | | $b$ ($b'$) | $v$ | $w$ | $x$ |
|---|---|---|---|---|---|
| $\boldsymbol{F_n}$ | | **1.1332** | **1.1316** | **1.1363** | **1.1477** |
| Errors in $F_n$ | $\sigma_A$ | 0.0085 | | | |
| | $\sigma_B$ | 0.0128 | 0.0122 | 0.0148 | 0.0161 |
| | $\sigma_C$ | 0.0042 | 0.0027 | 0.0038 | 0.0027 |
| **Total uncertainty in $\boldsymbol{F_n}$**† | | **0.0160** | **0.0152** | **0.0175** | **0.0184** |

Bias correction factor normalized at *v* band ($\widehat{F_n}$)

| Bands ($n$) | | $b$ ($b'$) | $v$ | $w$ | $x$ |
|---|---|---|---|---|---|
| $\boldsymbol{\widehat{F_n}}$ | | **1.0014** | **1** | **1.0041** | **1.0142** |
| Errors in $\widehat{F_n}$ | $\sigma_A$ | 0.0070 | 0 | 0.0070 | 0.0070 |
| | $\sigma_B$ | 0.0068 | 0 | 0.0047 | 0.0044 |

| $\sigma_C$ | 0.0044 | 0 | 0.0041 | 0.0034 |
|---|---|---|---|---|
| **Total uncertainty in $\widehat{F_n}$†** | **0.0107** | **0** | **0.0094** | **0.0089** |

† Total uncertainties are the root sum of squares of $\boldsymbol{\sigma_A}$, $\boldsymbol{\sigma_B}$, and $\boldsymbol{\sigma_C}$ (section 4.2).

## 7. Absolute albedo of Ryugu and Bennu after cross calibration

We update the geometric albedos of Ryugu and Bennu by applying the results of our cross calibration. We validate our newly obtained geometric albedos by comparing it with those observed by ground-based telescopes and the OSIRIS-REx Visible and InfraRed Spectrometer (OVIRS).

Ground-based telescopes derived the geometric albedo of both Ryugu and Bennu before spacecraft arrivals. Telescopes observed reduced magnitude ($H_V(\alpha)$), which were then fitted with photometric functions (e.g., IAU H-V formalism) to derive absolute magnitude ($H_V = H_V(0°)$). Geometric albedo ($p_V$) were estimated by correcting $H_V$ for the asteroid size using the following equation:

$$p_V = \left(\frac{D_0}{D_{eff}}\right)^2 10^{-0.4H_V}. \quad (8)$$

Here, $D_0$ is a constant of 1329 km (Fowler & Chillemi, 1992) and $D_{eff}$ is the effective diameter of the asteroid, i.e., $\pi\left(D_{eff}/2\right)^2$ is equal to the cross section of the asteroid. Before the arrivals of Hayabusa2 and OSIRIS-REx at the asteroids, uncertainties in the ground-based estimation of $D_{eff}$, which were on the order of 10 m (Müller et al., 2017; Emery et al., 2014), resulted in a >10% uncertainty in $p_V$. Subsequent observations by the two missions provided accurate shape models of the asteroids, significantly reducing the uncertainties down to 1 m to cm. Thus, we recalculated $p_V$ by combining $H_V$ measured with ground-based telescopes and $D_{eff}$ updated by spacecraft observations. The updated $D_{eff}$ were 906 m for Ryugu and 490 m for Bennu; these values were calculated by averaging the projected area of their shape models (stereo-photoclinometry-based model version 20 Mar 2020 for Ryugu and altimeter-based model version 21 for Bennu) over one full rotation. Fig. 12a shows the recalculated $p_V$. The recalculation with the updated $D_{eff}$ lowered the $p_V$ of Ryugu by 8% than those originally reported in Ishiguro et al. (2014). All three plots of Ryugu and four plots of Bennu are derived from the same dataset of $H_V(\alpha)$ but fitted with different photometric functions. Given that the dataset of $H_V(\alpha)$ is compiled from years of observations by various instruments, the accuracies of photometric function fitting, which are shown as error bars in Fig. 12, likely reflects the absolute radiometric uncertainty of ground-based telescope observations. The values of $p_V$ derived from different photometric functions exhibit a larger variation for Bennu because $H_V$ is extrapolated from non-opposition ($\alpha$>15°) data, whereas the variation is smaller for Ryugu because $H_V$ is more directly estimated from data observed near opposition ($\alpha$>0.3º).

ONC-T and MapCam also measured $p_V$ by intensively observing the asteroids with phase angles near 0º (Fig. 12b). We obtained the cross-calibrated $p_V$ of Ryugu and Bennu by applying the $v$-band value of $F_n$ obtained in this study to the pre-cross-calibrated $p_V$ of Bennu (Golish et al., 2020b) and preserving those of Ryugu (Tatsumi et al., 2020; Yokota et al., 2021). The *absolute* values of these cross-calibrated $p_V$ are accurate within the 3% uncertainty of the ONC-T's original radiometric calibration (section 3), which are shown as error bars in Fig. 12b. The cross-calibrated $p_V$ of Ryugu and Bennu are 4.1 ± 0.1% and 4.9 ± 0.1%, respectively.

The $p_V$ of Bennu derived from OVIRS data is 4.9 ± 0.3% (Zou et al., 2021; Fig. 12b). The error shows the accuracies in absolute radiometric calibration and photometric correction. This measurement is independent of MapCam-based $p_V$ because the targets used for their radiometric calibrations are different; OVIRS used the Earth and Bennu for its radiometric calibration (Simon et al., 2018; 2021).

By applying the results of cross calibration, the MapCam-based $p_V$ of Bennu became consistent with OVIRS and all ground-based telescope data within their errors, whereas that before cross calibration show a larger discrepancy (Fig. 12). For instance, the 15% discrepancy between OVIRS and MapCam observed before cross calibration (as also pointed out by Golish et al., 2022) reduced to 1.6% after applying our results. The only apparent exception to this was the ground-based $p_V$ derived from linear fitting, which has been adjusted for the opposition effect assuming an opposition amplitude (i.e., deviation from the linear fit at $\alpha$ = 0º) of $0.05^{+0.15}_{-0.05}$ mag (Hergenrother et al., 2013); this $p_V$ exhibited better agreement with the pre-cross-calibrated value. Although the opposition amplitude had to be assumed in the original report due to the unavailability of near-opposition data, detailed phase curve observations by spacecraft have later shown that the opposition amplitude of Bennu is 0.20 mag (Hergenrother et al., 2019). After updating the opposition amplitude to the actual observed value, the $p_V$ became more consistent with the cross-calibrated data, further supporting the reliability of our result (Fig. 12). These overall consistencies verify that the absolute radiometric accuracy of MapCam data improved to 3% after cross calibration.

The $p_V$ of Ryugu observed by ONC-T are also consistent with the ground-based values within their errors, except for that derived from the H-G function (Fig. 12). This discrepancy may be attributed to systematic model errors inherent in the H-G function, which has been argued to overestimate the absolute magnitude when fitted across a wide range of $\alpha$ (Ishiguro et al., 2014; Hergenrother et al., 2019). We also observe that relying on the original calibration of ONC-T and upscaling the MapCam data (our current approach; equation 4) results in the best match between the ground-based and spacecraft observations; relying on MapCam and downscaling the ONC-T data would lead to a larger discrepancy.

Furthermore, we confirm that applying the results of our cross calibration does not significantly alter the band ratios of Ryugu and Bennu, ensuring that their spectral shapes remain consistent with ground-based telescope observations (Fig. 13).

(a) Ground-based telescope observations

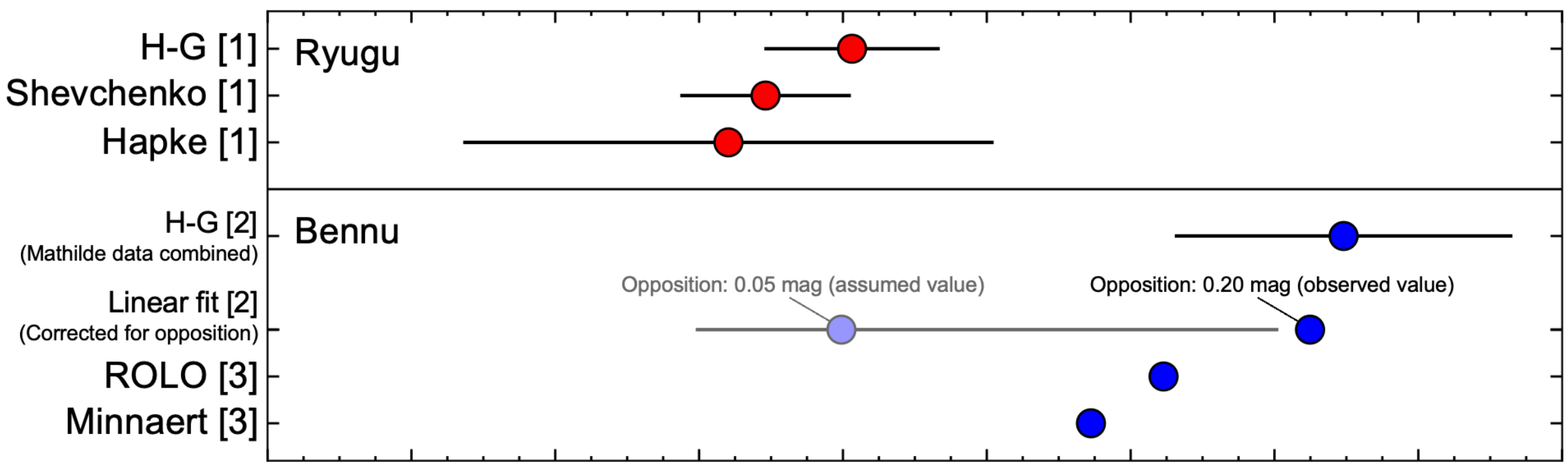

(b) Spacecraft observations

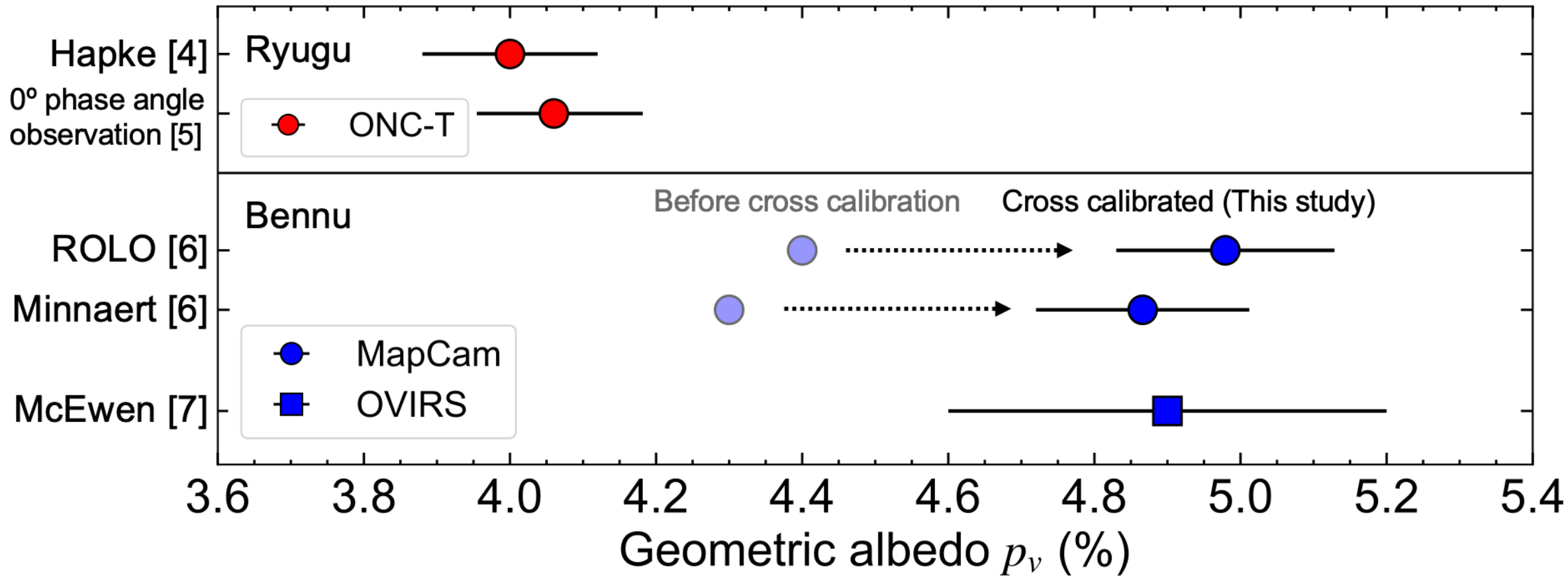

**Figure 12.** Geometric albedo ($\boldsymbol{p_v}$) of Ryugu and Bennu observed by **(a)** ground-based telescopes and **(b)** spacecraft. Labels in the vertical axis show the photometric functions used for derivation. The error bars in (a) show the accuracy in fitting the observation data with photometric functions, where applicable. Error bars for the ONC-T and MapCam data in (b) show their 3% absolute radiometric accuracy after cross calibration. Error bars for the OVIRS data show accuracies in absolute radiometric calibration and photometric correction. [1] Ishiguro et al. (2014). [2] Hergenrother et al. (2013). [3] Takir et al. (2015). [4] Tatsumi et al. (2020). [5] Yokota et al. (2021). [6] Golish et al. (2020b). [7] Zou et al. (2021).

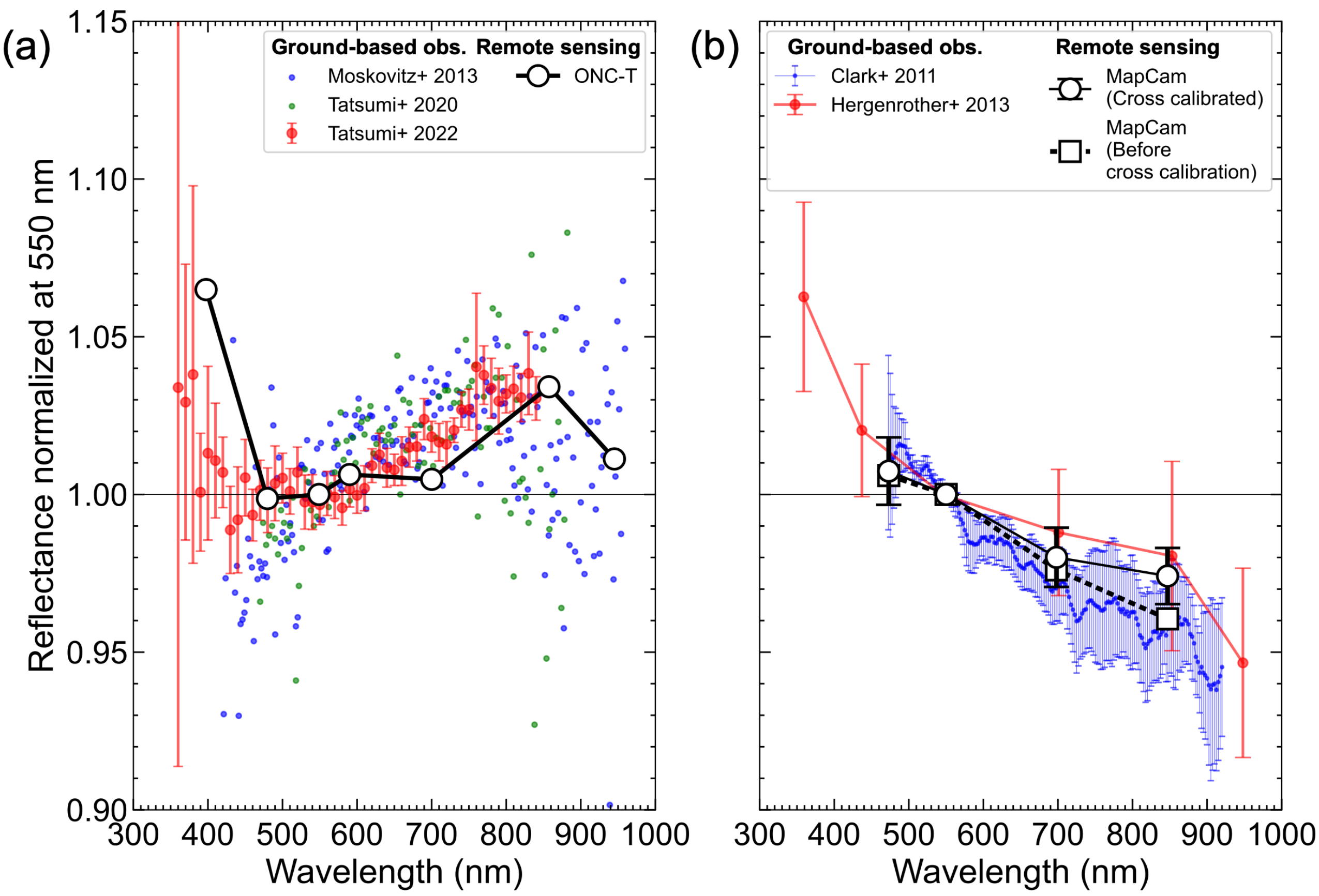

**Figure 13.** Spectra of (**a**) Ryugu and (**b**) Bennu compared with ground-based telescope observations. Error bars of the cross-calibrated spectrum of Bennu show the uncertainty in cross calibration (i.e., uncertainty in $\widehat{\boldsymbol{F}_n}$).

## 8. Conclusion

We conducted a cross calibration between the multiband imagers ONC-T on Hayabusa2 and MapCam on OSIRIS-REx to compare the spectra of Ryugu and Bennu at the shared *b*, *v*, *w*, and *x* bands (0.48−0.85 μm) with improved accuracy. Such quantitative comparison was not possible before because the two cameras were calibrated with different targets; ONC-T used stars while MapCam used the Moon. In this study, we used the Moon as the common standard to correct the imager-to-imager bias. Since the two imagers observed different faces of the Moon with different illumination/viewing geometries, we compared the observed lunar images against those simulated by two lunar photometry models derived from Kaguya's SP and LRO's WAC data. We carefully evaluated their limitations and errors, especially in the high phase angle range of 42°–59°.

We derived bias correction factors ($F_n$) for each of the $n = b$, *v*, *w*, and *x* bands (Table 5), which compensate for the bias of MapCam to ONC-T caused by their difference in solar irradiance models used for data reduction and targets used for radiometric calibrations. Future studies can simply obtain the cross-calibrated reflectance of Ryugu and Bennu by multiplying the pre-cross-calibrated reflectance data of Bennu by $F_n$ while retaining the reflectance of Ryugu. These pre-cross-calibrated data are publicly available via the Planetary Data System. The uncertainties in $F_n$ constrain the *relative difference* between the cross-calibrated data of Ryugu and Bennu. Since $F_n$ corrects the bias of MapCam to ONC-T, the *absolute* reflectances after cross calibration are accurate within the 3% uncertainty of ONC-T's original radiometric calibration based on observations of standard stars.

Using the cross-calibrated data is crucial for the quantitative comparison of Ryugu and Bennu data because our cross calibration revealed a significant imager-to-imager bias between ONC-T and MapCam. For instance, we showed that the pre-cross-calibrated reflectance of Bennu needs to be upscaled by 13.2 ± 1.5% at *v* band to correct the imager-to-imager bias; the uncertainty shows that we can compare the reflectances of Ryugu and Bennu with an improved precision of 1.5% after applying our cross calibration. In contrast, we showed that the imager-to-imager bias in band ratio was smaller than 1.5%. The cross-calibration uncertainties in the band ratios were 0.9−1.1% (or ±0.03 $\mu m^{-1}$ in *b*-to-*x* band spectral slope).

The cross-calibrated geometric albedo of Ryugu was 4.1 ± 0.1% and that of Bennu was 4.9 ± 0.1%. By applying our cross calibration, these albedos became consistent with those observed by ground-based telescopes and the OSIRIS-REx Visible and InfraRed Spectrometer (OVIRS), supporting the reliability of our result.

**Acknowledgements**

We are grateful to the entire OSIRIS-REx and Hayabusa2 teams for making the encounters with Bennu and Ryugu possible. The LROC/LRO-L-LROC-5-RDR-V1.0 dataset was obtained from the Planetary Data System (PDS). K.Y acknowledges funding from JSPS Fellowship (Grant number JP21J20894) and International Graduate Program for Excellence in Earth-Space Science (IGPEES) from the University of Tokyo. E. T. and S. S acknowledge funding by International Visibility Program of Hayabusa2# project from JAXA. This project has been supported by JSPS grants (Grant numbers 22K21344 and 23H00141).

**Data Availability Statement**

The calibrated (L1, iofL2) MapCam images are available via the Planetary Data System (PDS): https://sbn.psi.edu/pds/resource/orex/. The calibrated (L2b, L2d) ONC-T images are also available via PDS: https://sbn.psi.edu/pds/resource/hayabusa2/.

**Supplementary materials**

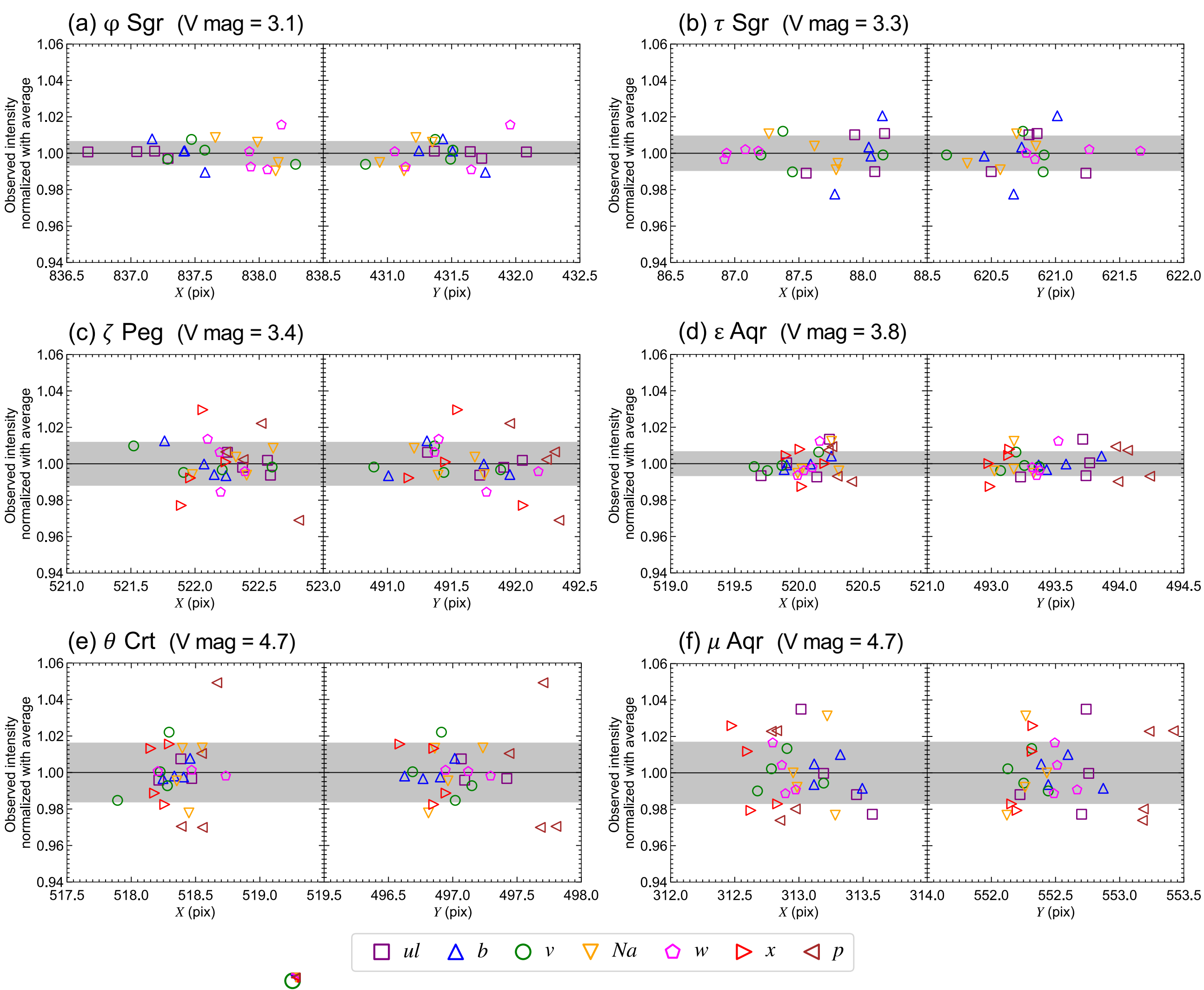

**Figure S1.** Precisions in the ONC-T observations of the other six darker standard stars are depicted using the same symbols and notations as in Fig. 5.

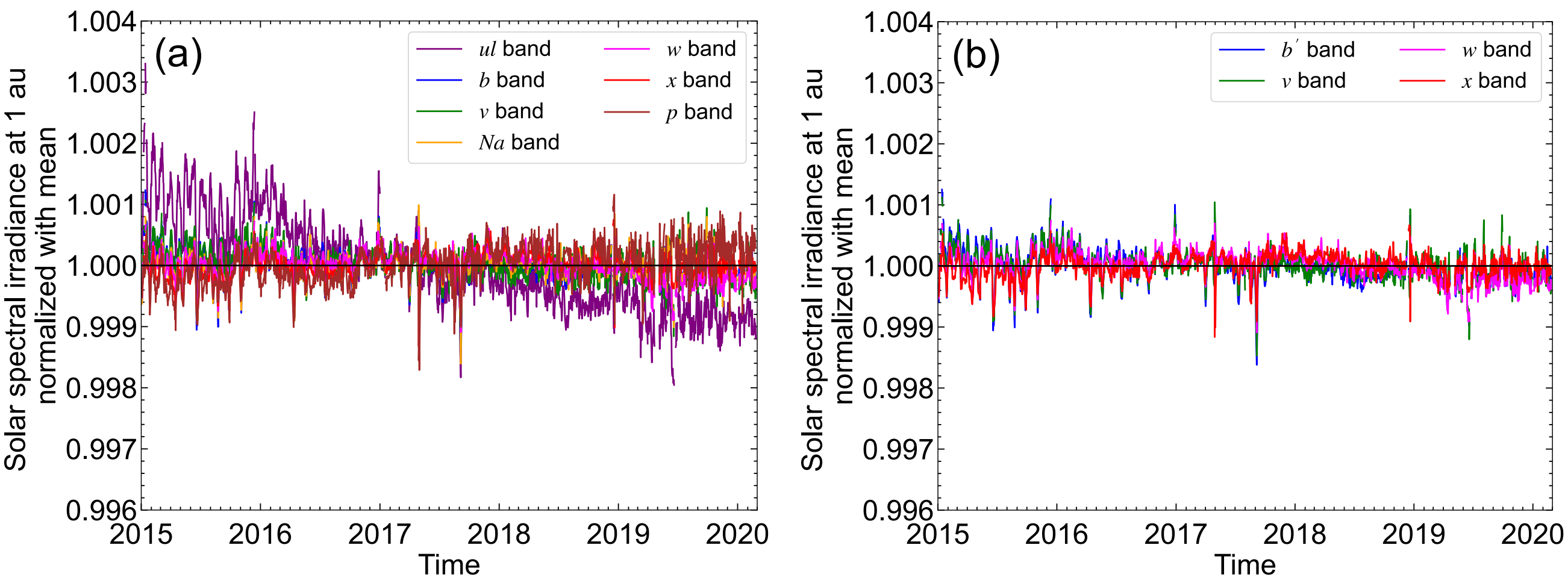

**Figure S2.** Time variations of solar spectral irradiance for **(a)** ONC-T and **(b)** MapCam filters calculated from daily data by the SORCE satellite (Harder et al., 2020).

## A. Detailed procedure for lunar image simulation

We first prepared lunar reflectance maps photometrically normalized to a standard illumination/viewing condition: $r_{norm}(i_{std}, e_{std}, \alpha_{std}, \lambda)$. The $r_{norm}$ for the WAC model was developed by Sato et al. (2014) and that for the SP model was developed by Kouyama et al. (2016). The standard illumination/viewing conditions are $(i_{std}, e_{std}, \alpha_{std})$ = (30º, 0º, 30º) for the SP model and $(i_{std}, e_{std}, \alpha_{std})$ = (60º, 0º, 60º) for the WAC model. We resampled the resolution of $r_{norm}$ to 0.5ºN × 0.5ºE to unify the spatial resolutions.

We note that $r_{norm}$ of WAC and SP models are highly consistent with each other (Fig. S3) because radiometric calibrations of both models used the data observed by the Robotic Lunar Observatory (ROLO; Kieffer & Stone, 2005). For instance, Mahanti et al. (2016) updated the RCCs of WAC images to achieve consistency with the ROLO data. Similarly, Kouyama et al. (2016) applied the following function ($p$) to the spectra observed by SP to match the observations by ROLO:

$$p(\lambda) = a_0 + a_1\lambda + a_2\lambda^2 + a_3\lambda^3. \quad (9)$$

The coefficients were $a_0 = 2.185$, $a_1 = -2.764 \times 10^{-3}$, $a_2 = 2.026 \times 10^{-6}$, and $a_3 = -5.056 \times 10^{-10}$; we updated these parameters from Kouyama et al. (2016) in this study to achieve better matching accuracy with ROLO.

To project $r_{norm}$ to the image plane, we calculated the latitude/longitude and illumination/viewing geometries for each of the observed image with an eight times higher spatial resolution ("Observation Geometries" in Fig. 7). We calculated these geometries by using the spacecraft position and attitude data stored in the SPICE kernels (Acton, 1996) and assuming a spherical shape of the Moon. We further refined the spacecraft attitude using the observed lunar images (Ogohara et al., 2012) to obtain accuracy below the plate scale (i.e., <1 pix error). We used the calculated latitude/longitude geometries to project $r_{norm}$ to the image plane. This results in a hyperspectral image cube of $r_{norm}$ with eight times higher resolution than the observed image (Fig. 7a).

To simulate the reflectance observed under an illumination/viewing geometry of $(i, e, \alpha)$, we photometrically corrected $r_{norm}$ using the following equation:

$$\tilde{r}_{sim}(i, e, \alpha, \lambda) = r_{norm}(i_{std}, e_{std}, \alpha_{std}, \lambda) \cdot \frac{r_{model}(i, e, \alpha, \lambda)}{r_{model}(i_{std}, e_{std}, \alpha_{std}, \lambda)}. \quad (10)$$

Here, $r_{model}$ is the reflectance modelled using the WAC and SP models, and $\tilde{r}_{sim}$ is the simulated reflectance. This photometric correction results in a hyperspectral image cube of $\tilde{r}_{sim}$ (Fig. 7b). In our calculation of equation 10, we corrected the phase function of the SP model to match that of the WAC model. Since ONC-T and MapCam observed the Moon with phase angles ($\alpha$) of 59º and 42º, respectively (Table 4), using $r_{model}$ with accurate phase functions (i.e., accurate $r_{model}(\alpha = 59°)/ r_{model}(\alpha = 42°)$) is important to obtain unbiased $\tilde{r}_{sim}$. We found that the phase functions of the WAC and SP models have a systematic difference of up to 3% (Fig. S4). The phase function of the WAC model is likely more accurate at phase angles near 60º because it is derived using higher phase angle data ($\alpha$<97º) than the SP model ($\alpha$<75º). In addition, the phase function of the Moon depends strongly on geologic unit (e.g., Helfenstein & Veverka, 1987), and the WAC model resolves such local effects at much higher resolution. Thus, we applied the following correction function to $\tilde{r}_{sim}$ of the SP model for all bands to resolve the systematic model-to-model difference:

$$q(\alpha) = b_0 + b_1\alpha + b_2\alpha^2. \quad (11)$$

Here, the phase angle $\alpha$ is given in degrees and the fitted coefficients were $b_0 = 8.992 \times 10^{-1}$, $b_1 = 5.069 \times 10^{-3}$, and $b_2 = -6.470 \times 10^{-5}$.

We calculated the in-band average of $\tilde{r}_{sim}$ for each of the filters on ONC-T and MapCam ($r_{sim,n}$) using the following equation:

$$r_{sim,n} = \frac{\int \tilde{r}_{sim}(\lambda)\, \lambda\phi_n(\lambda)\, d\lambda}{\int \lambda\phi_n(\lambda)\, d\lambda}. \quad (12)$$

Since the wavelengths defined in the WAC model are sparser than the band widths of filters on ONC-T and MapCam, we increased the wavelength resolution of $\tilde{r}_{sim}$ to 1 nm using linear interpolation prior to computing equation 12. This approach is reasonable because the lunar spectra is linear in the 400–600 nm range (Pieters, 1999). In addition, we populated regions with data deficiency (e.g., regions near the poles) with the global average reflectance. After these procedures, we obtain $r_{sim,n}$ with eight times higher resolution than the observed image (Fig. 7c).

After optimizing the co-registration with the observed image (appendix B), we downscaled $r_{sim,n}$ to match the observed resolution (i.e., 8 pix × 8 pix binning) and convolved it with a Gaussian-approximated Point Spread Function (PSF) to simulate the optical modulation. The PSF width of ONC-T was determined from inflight observations of stars (Suzuki et al., 2018). For MapCam, line-spread functions (LSF), which are 1-D integral representations of the 2-D PSFs, were measured before launch (Rizk et al., 2018). We utilized the LSF width measured in the worst-performing one-dimensional axis as the PSF width for simplicity. Following these procedures, we obtained images of $r_{sim,n}$ with the same resolution as the observed image (Fig. 7d).

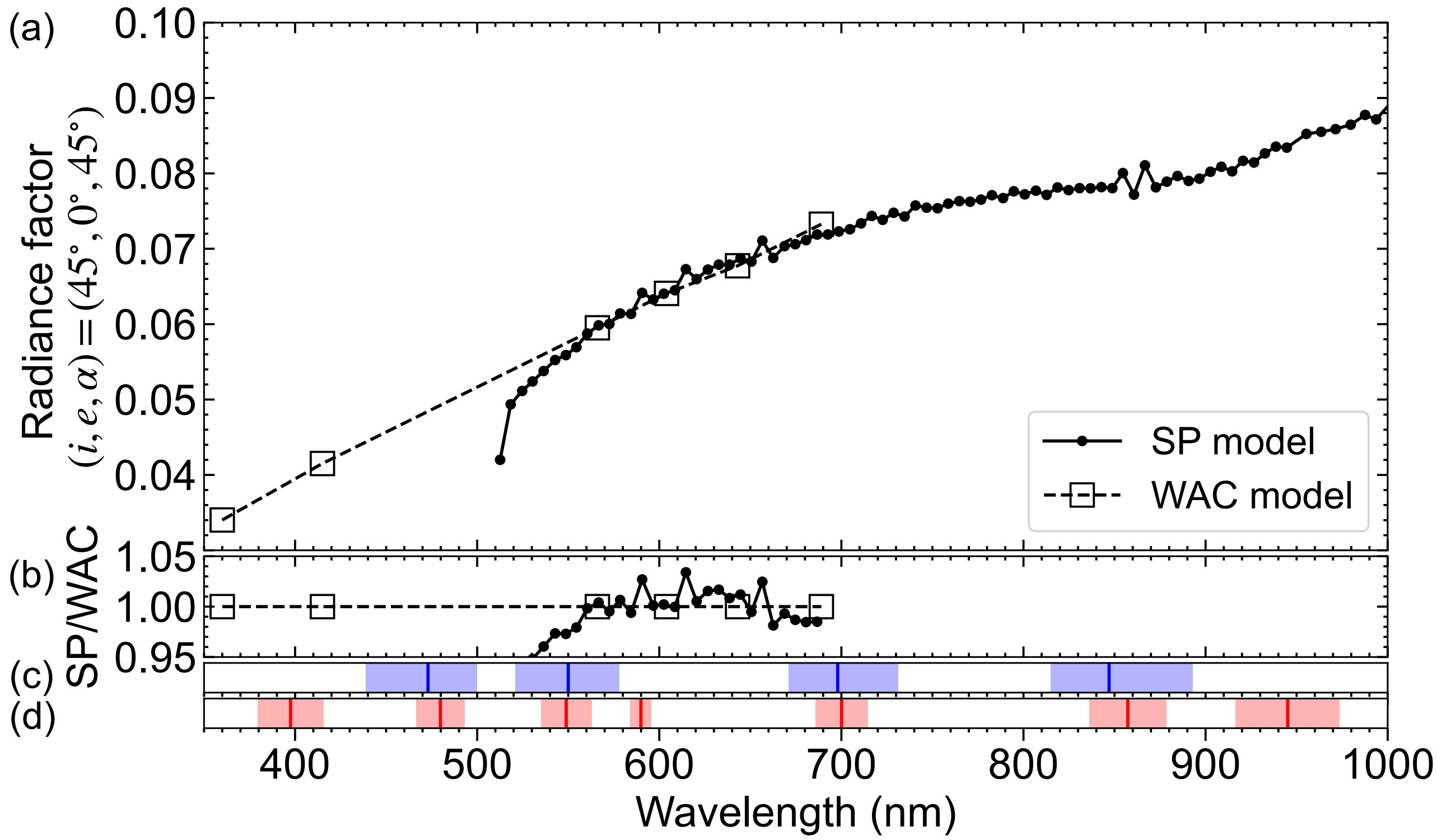

**Figure S3.** **(a)** Globally (70ºS – 70ºN) averaged lunar reflectance of the WAC and SP models photometrically normalized to a standard condition of ( $i_{std}, e_{std}, \alpha_{std}$ ) = (45º, 0º, 45º): $r_{norm}(45°, 0°, 45°, \lambda)$. **(b)** The ratio of the SP-model spectrum to the WAC-model spectrum. **(c)** The effective band centers (solid vertical lines) and the cut-on/cut-off wavelengths (hatches) of all four filters onboard MapCam (Golish et al., 2020a). **(d)** The effective band centers (solid vertical lines) and effective band widths (hatches) of all seven filters onboard ONC-T (Tatsumi et al., 2019).

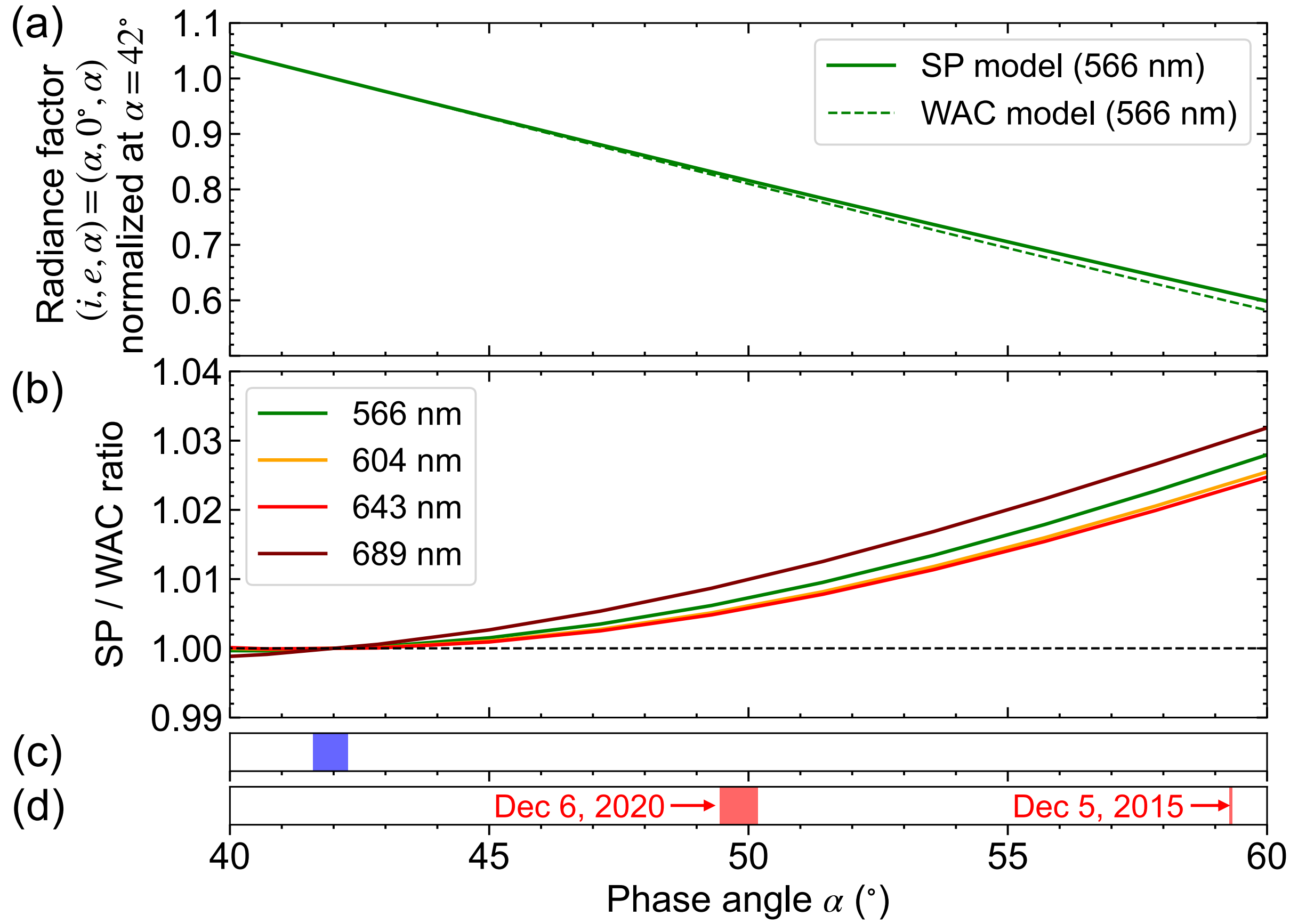

**Figure S4. (a)** Globally (70ºS – 70ºN) averaged phase functions of the WAC and SP models at $\lambda$ =566 nm: $r_{model}(\alpha, 0^\circ, \alpha, 566\ \text{nm})$. **(b)** The ratio of the phase functions between the SP and WAC models at 566, 604, 643, and 689 nm. The phase angle ranges during the lunar observations by MapCam and ONC-T are shown in **(c)** and **(d)**, respectively.

## B. Co-registration of the simulated lunar image with the observed image

The simulated image ($r_{sim,n}$) needs to be accurately co-registered to the observed image ($r_{obs,n}$) to perform their pixel-by-pixel comparison. We needed to conduct such a pixel-by-pixel comparison because certain pixels of $r_{sim,n}$ have large errors as discussed in section 4.1.2, and we must exclude them to obtain accurate $\overline{r_{sim,n}/r_{obs,n}}$.

We carried out fine co-registration between $r_{sim,n}$ and $r_{obs,n}$ by optimizing the image shifts in the vertical/horizontal directions ($\Delta X, \Delta Y$; in units of pixels in the observed image) and the rotation angle ($\Delta\theta$) with respect to the center of the lunar disk. We searched for the best ($\Delta X, \Delta Y, \Delta\theta$) combination which gives the highest $r_{sim,n}$-to-$r_{obs,n}$ correlation using grid search. We varied ($\Delta X$, $\Delta Y$) in the range of -1.5 to +1.5 pix with 0.025 pix increment and $\Delta\theta$ in the range of -2º to +2º with 0.1º increment.

We determined the required co-registration accuracies based on numerical experimentation. We calculated the disk-averaged ratio between the two identical $r_{sim,n}$ shown in Fig. 7d with either one intentionally shifted by $(\Delta X,\ \Delta Y)$ and rotated by $\Delta\theta$; the intentional shifts and rotation represent the mis-registered condition. Fig. S5 shows the disk-averaged ratio and pixel-by-pixel correlation as a function of $(\Delta X, \Delta Y, \Delta\theta)$. Since these two images are identical, the ratio and correlation are unity at $(\Delta X, \Delta Y, \Delta\theta)$ = (0, 0, 0). However, misregistration of one image to the other by $(\Delta X, \Delta Y, \Delta\theta)$ results in a deviation from unity, causing systematic errors in their ratio. The ratio is most sensitive to $\Delta X$ in the case of Fig. 7d because the reflectance variation is largest along this axis. Thus, subtle misalignment of $\Delta X$ by only 0.5 pix causes a 2% error in the disk-averaged ratio. From such analyses, we concluded that $(\Delta X,\ \Delta Y)$ needs to be determined with an accuracy of <0.2 pix and $\Delta\theta$ needs to be determined with <10º to achieve <1% accuracy in $\overline{r_{sim,n}/r_{obs,n}}$.

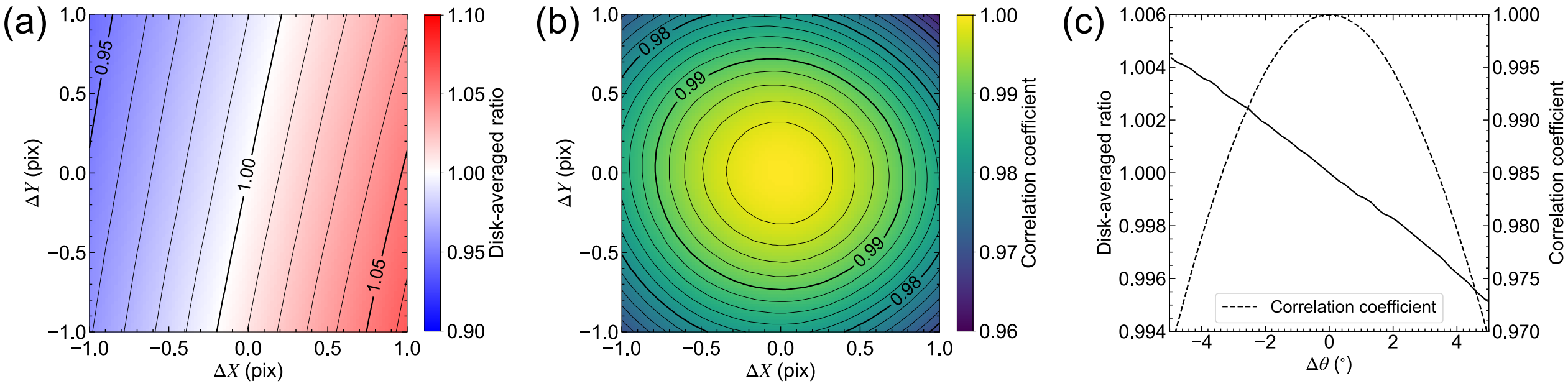

**Figure S5.** Simulation results of the image-to-image misregistration effect. We calculated the ratio and correlation of two identical lunar images (Fig. 7d) with registration errors of $(\Delta X,\ \Delta Y)$ in image shifts (pix) and $\Delta\theta$ (°) in rotation angle. **(a)** The ratio as a function of $(\Delta X,\ \Delta Y)$. **(b)** The correlation as a function of $(\Delta X,\ \Delta Y)$. **(c)** The ratio (solid curve) and correlation (dashed curve) as a function of $\boldsymbol{\Delta\theta}$.

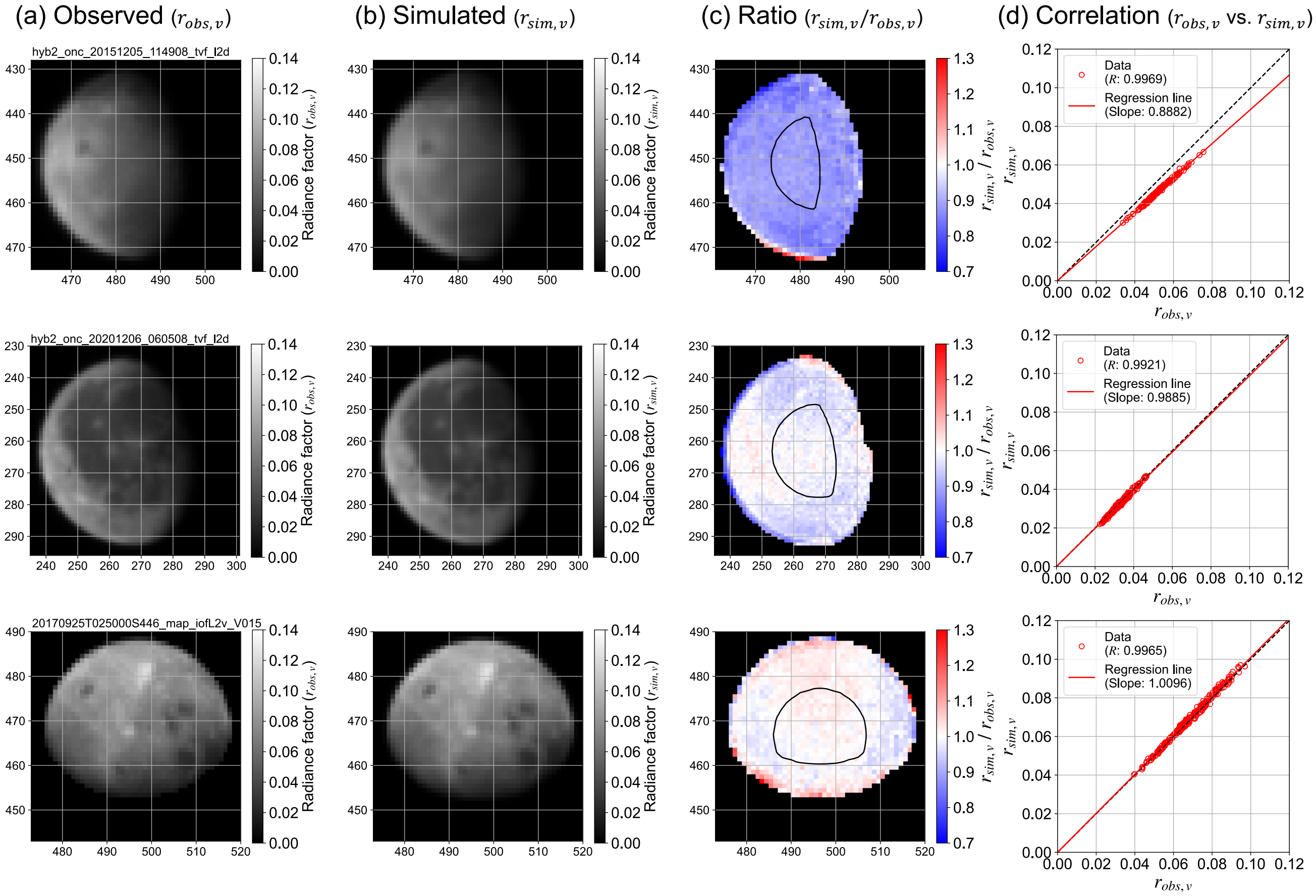

**Figure S6.** Comparison between **(a)** the observed image ($r_{obs,n}$) and **(b)** image simulated using the SP model ($r_{sim,n}$) for the **(top row)** ONC-T observations after launch (Dec 5, 2015**), (middle row)** ONC-T observations during the returning cruise (Dec 6, 2020), and **(bottom row)** MapCam observations (Sep 25, 2017). The ratio image ($r_{sim,n}/r_{obs,n}$) is shown in **(c)**. Pixels within the dark curve (incidence angle <60º, emission angles <30º, and latitudes <70ºS – 70N) were averaged. Pixel-by-pixel correlation between $r_{obs,n}$ and $r_{sim,n}$ within the dark curve is shown in **(d)**. The same plots with $r_{sim,n}$ based on the SP model are shown in Fig. S6.